\documentclass[a4paper,11pt]{article}
\pdfoutput=1 

\usepackage{jheppub}
\usepackage{url}
\usepackage{subcaption}
\usepackage{rotating}

\usepackage[T1]{fontenc} 

\title{\boldmath Precision measurements on $\delta_\text{CP}$ in MOMENT}

\author[]{Jian Tang,}
\author[]{Sampsa Vihonen,}
\author[]{Tse-Chun Wang}

\affiliation[]{School of Physics, Sun Yat-Sen University,\\
No. 135, Xingang Xi Road, Guangzhou, 510275, P. R. China}

\emailAdd{tangjian5@mail.sysu.edu.cn}
\emailAdd{sampsa@mail.sysu.edu.cn}
\emailAdd{wangzejun@mail.sysu.edu.cn}

\abstract{As it is very promising to expect a discovery of CP violation in the leptonic sector, the precision measurement of the Dirac CP phase $\delta_\text{CP}$ is going to be one of the key interests in the future neutrino oscillation experiments. In this work, we examine the physics reach of the proposed medium baseline muon decay experiment MOMENT. In order to identify potential bottlenecks and opportunities to improve CP precision in MOMENT, we investigate the effect of statistical error, systematic uncertainties, fraction of the muon beam polarity, and adjusting the baseline length to match the first or second oscillation maximum on the precision measurement of $\delta_\text{CP}$. We also simulate superbeam experiments T2K, NO$\nu$A, T2HK, DUNE and T2HKK in comparison and complementary to MOMENT. To reach the precision of $\delta_\text{CP}$ at 12$^\circ$ or better at 1$\,\sigma$ confidence level, we find it sufficient to combine the data of MOMENT, DUNE and T2HK. }

\begin{document} 
\maketitle
\flushbottom

\section{Introduction}
\label{Section:1}

Neutrino physics has experienced a steady flow of progress since the confirmation of neutrino oscillations was made in Super-Kamiokande and Sudbury Neutrino Observatory nearly two decades ago~\cite{Kajita:2000mr,Ahmad:2001an}. This has been achieved in the many reactor, solar, atmospheric and accelerator neutrino experiments that have operated since then.

The oscillation of three neutrinos can be parametrized in terms of six physical variables~\cite{Pontecorvo:1957cp,Maki:1960ut,Kobayashi:1973fv}: the three mixing angles $\theta_{12}$, $\theta_{13}$ and $\theta_{23}$, the two mass-squared differences $\Delta m_{21}^2=m_2^2-m_1^2$ and $\Delta m_{31}^2=m_3^2-m_1^2$, and the Dirac CP phase $\delta_\text{CP}$. These parameters have been measured to a high precision (see Refs.~\cite{NuFit:4-1,Esteban:2018azc,deSalas:2018bym,Capozzi:2018ubv} for reviews). Whereas the best-fit values of the so called solar parameters $\theta_{12}$ and $\Delta m_{21}^2$ have been determined with the relative precisions of 14\% and 16\% at 3$\,\sigma$ confidence level (CL), respectively, the precision of the so called reactor mixing angle $\theta_{13}$ is 8.9\%~\cite{NuFit:4-1}. The value of the atmospheric mixing angle $\theta_{23}$ is less precisely known, with 23\% at 3$\,\sigma$, and it is not well known whether it resides in the high octant $\theta_{23} >$~45$^\circ$, or the low octant, $\theta_{23} <$~45$^\circ$. The sign of the atmospheric mass splitting $\Delta m_{3\ell}^2$, where $\ell =$ 1 for the normal ordering ($m_1 < m_2 \ll m_3$) and 2 for inverted ordering ($m_3 \ll m_1 < m_2$), is currently unknown, but its magnitude $\left| \Delta m_{3\ell}^2 \right|$ has been measured with the relative precision of 7.5\%. Most recently, a global fit from the neutrino oscillation data has provided a best-fit for the Dirac CP phase at $\delta_\text{CP} \simeq$ 221$^\circ$, when the mass ordering is normal, and $\delta_\text{CP} \simeq$ 282$^\circ$ when the ordering is inverted~\cite{NuFit:4-1}. The signal remains statistically insignificant, however, and the values of $\delta_\text{CP} =$ 144...357$^\circ$ are still available within the 3$\,\sigma$ CL bounds.

Precision measurements on the neutrino oscillation parameters open up an opportunity to search for the origin of neutrino mixing, e.g. a flavour symmetry~\cite{King:2015aea}. One of the key parameters in discriminating between the various available flavour symmetries is the Dirac CP phase $\delta_\text{CP}$, which is desirable to be known by approximately 10$^\circ$ uncertainty at 1$\,\sigma$ CL in the event where the value $\delta_\text{CP}$ is confirmed to lie near 270$^\circ$ (see e.g. Ref.~\cite{Petcov:2018snn}). The precision measurements of $\delta_\text{CP}$ in future experiments have been previously discussed in e.g. Refs.~\cite{Ballett:2016daj,Ghosh:2019sfi,Coloma:2012wq,Winter:2003ye,Smirnov:2018ywm}. 

Future neutrino oscillation experiments will play an important role in curtailing the error on $\delta_\text{CP}$. In this work, we investigate the precision measurement of $\delta_\text{CP}$ in the MuOn-decay MEdium baseline NeuTrino beam facility (MOMENT)~\cite{Cao:2014bea}. MOMENT is proposed to use a novel approach to generate a very intensive beam of neutrinos and antineutrinos via the decay of positive and negative muons, which allows to survey neutrino oscillations in eight different channels. In the past, MOMENT and its physics potential have been studied in CP violation discovery~\cite{Blennow:2015cmn,Bakhti:2016prn}, non-standard neutrino interactions~\cite{Bakhti:2016prn,Tang:2017qen} and invisible neutrino decays~\cite{Tang:2018rer}.

In the present work, we study the physics potential of MOMENT in the standard three neutrino oscillations and compare its sensitivities with those of the currently running long baseline neutrino experiments T2K~\cite{Abe:2011ks} and NO$\nu$A~\cite{Ayres:2007tu}, as well as the next generation experiments T2HK~\cite{Abe:2015zbg} and DUNE~\cite{Acciarri:2015uup}. We also discuss the option of placing the second Hyper-Kamiokande detector in Korea, as was suggested in the T2HKK~\cite{Abe:2016ero} proposal. We find that MOMENT has an excellent sensitivity for the precision measurement of $\delta_\text{CP}$ at its current best-fit values. We also find that MOMENT can measure $\delta_\text{CP}$ by 15$^\circ$ uncertainty or better at 1$\,\sigma$ CL for approximately 76\% of its currently allowed values. 

This paper is organized as follows: In Section~\ref{Section:2}, we review the oscillation probabilities relevant for MOMENT and discuss the role of the first and second oscillation maxima in the precision measurement of $\delta_\text{CP}$. In Section~\ref{Section:3}, we give a brief introduction to the experimental setup of MOMENT and describe how its simulation is done in this work. We present the results of this work in Section~\ref{Section:4}, and summarize our findings in Section~\ref{Section:6}.

\section{A brief review of the oscillation probabilities in CP precision}
\label{Section:2}

In this section, we perform an analytical study on the neutrino oscillation probabilities that are relevant for $\delta_\text{CP}$ precision in the experiments considered in this work.

The precision on the $\delta_\text{CP}$ parameter can be extracted in neutrino oscillation experiments by comparing the spectral data from the neutrino and antineutrino oscillation channels. In conventional beam experiments such as T2K and NO$\nu$A, the sensitivity to the $\delta_\text{CP}$ parameter is mainly facilitated by comparing the measured event and energy spectra in the appearance channels $\nu_\mu \rightarrow \nu_e$ and $\bar{\nu}_\mu \rightarrow \bar{\nu}_e$, from which the value of $\delta_\text{CP}$ can then be resolved.

Neutrino experiments based on muon decay are different from superbeams by the number of oscillation channels which can be used for the determination of the $\delta_\text{CP}$ value. MOMENT, for example, benefits from the simultaneous access to the channels $\nu_e \rightarrow \nu_\mu$ and $\bar{\nu}_\mu \rightarrow \bar{\nu}_e$ when the experiment runs in the $\mu^+$ beam mode, and to their conjugate channels $\bar{\nu}_e \rightarrow \bar{\nu}_\mu$ and $\nu_\mu \rightarrow \nu_e$ when it is in the $\mu^-$ mode. When the corresponding disappearance channels are taken into account, the experiment gains access to eight channels in total.

Let us begin by considering the oscillations of the neutrinos that originate from the $\mu^+$ beam. The probabilities for the appearance channels $\nu_e \rightarrow \nu_\mu$ and $\bar{\nu}_\mu \rightarrow \bar{\nu}_e$ can be approximated with the following analytical expressions~\cite{Ballett:2016daj,Coloma:2012wq,Asano:2011nj}. Taking the quantity $\varepsilon \equiv \Delta m_{21}^2 / \Delta m_{31}^2$ to be small, such that $\sin^2 \theta_{13} \sim \mathcal{O}(\varepsilon)$, the appearance probabilities can be expressed in powers of $\varepsilon$. For example, in the case of $\nu_e \rightarrow \nu_\mu$ channel, one gets
\begin{equation}
P_{\nu_e \rightarrow \nu_\mu} = P_1 + P_{3/2} + \mathcal{O}(\varepsilon^2).
\label{prob:01}
\end{equation}
Whereas the first-order term $P_1$ is independent of $\delta_\text{CP}$,
\begin{equation}
P_1 = \frac{4}{(1-r_A)^2} \, \sin^2 \theta_{23} \, \sin^2 \theta_{13} \, \sin^2 \left( \frac{(1-r_A) \, \Delta \, L}{2} \right),
\label{prob:02}
\end{equation}
the other term depends on the value of the $\delta_\text{CP}$ parameter:
\begin{equation}
P_{3/2} = 8 \, J_r \, \frac{\varepsilon}{r_A \, (1-r_A)} \, \cos \left( \delta_\text{CP} - \frac{\Delta \, L}{2} \right) \, \sin \left( \frac{r_A \, \Delta \, L}{2} \right) \, \sin \left( \frac{(1-r_A) \, \Delta \, L}{2} \right).
\label{prob:03}
\end{equation}
In the above expressions $J_r = \cos\theta_{12} \, \sin\theta_{12} \, \cos\theta_{23} \, \sin\theta_{23} \, \sin \theta_{13}$ is the reduced Jarlskog invariant, $L$ is the baseline length and $E$ is the energy of the neutrino or antineutrino. The matter effects enter the probability via $r_A = \sqrt{2} \, G_F \, N_e / \Delta$, where $N_e$ is the electron number density, $G_F$ the Fermi coupling constant and $\Delta \equiv \Delta m_{31}^2 / 2E$.

As is apparent in Eqs.~(\ref{prob:01}--\ref{prob:03}), the sensitivity to $\delta_\text{CP}$ arises from the $P_{3/2}$ term, which is dependent on the value of $\delta_\text{CP}$ through $\cos(\delta_\text{CP} - \Delta \, L /2)$. The maximum value of $P_{3/2}$ is therefore attained when $\Delta \, L /2 \simeq (2n+1) \, \pi/2$, $n =$ 0, 1, 2,..., and the minimum when $\Delta \, L /2 \simeq n \, \pi$, respectively. The first oscillation maximum is said to occur when $\Delta \, L /2 \simeq \pi/2$, and the second oscillation maximum when $\Delta \, L /2 \simeq 3\, \pi/2$. When the condition for either oscillation maximum is met, the dependence on $\delta_\text{CP}$ can be approximated as $P_{3/2} \propto \sin \delta_\text{CP}$.

The other probability relevant to the $\mu^+$ beam, $P_{\bar{\nu}_\mu \rightarrow \bar{\nu}_e}$, can be conveniently obtained from Eqs.~(\ref{prob:01}-\ref{prob:03}) via the transformation $r_A \rightarrow -r_A$. Knowing that the matter potentials are small in medium-baseline experiments like MOMENT, we remark that $P_{\nu_e \rightarrow \nu_\mu}$ and $P_{\bar{\nu}_\mu \rightarrow \bar{\nu}_e}$ take approximately the same size and shape.

When the oscillations originating from the $\mu^-$ beam are concerned, the relevant oscillation channels for $\delta_\text{CP}$ measurements are $\nu_\mu \rightarrow \nu_e$ and $\bar{\nu}_e \rightarrow \bar{\nu}_\mu$. The expression for the $\nu_\mu \rightarrow \nu_e$  probability can be acquired from Eqs.~(\ref{prob:01}--\ref{prob:03}) through the transformation $\delta_\text{CP} \rightarrow - \delta_\text{CP}$. The probability for $\bar{\nu}_e \rightarrow \bar{\nu}_\mu$ can be similarly obtained via the transformations $\delta_\text{CP} \rightarrow -\delta_\text{CP}$ and $r_A \rightarrow -r_A$, leading to similar behaviour near the oscillation maxima.

One can make an estimate on the size of the relative uncertainty that is attainable on the true value of $\delta_\text{CP}$ in a neutrino oscillation experiment~\cite{Ballett:2016daj,Coloma:2012wq}. Owing to the sinusoidal nature of the $\sin \delta_\text{CP}$ term in $P_{3/2}$, one can expect the best resolution to the individual values of $\delta_\text{CP}$ near the CP conserving values $\delta_\text{CP} =$ 0, $\pi$ and the worst resolution near $\delta_\text{CP} = \pi/2$, $3\pi/2$, respectively. When the oscillations take place in vacuum ($r_A =$ 0) the correspondence is exact, whereas the presence of matter effects shifts the maximal and minimal sensitivities away from their respective values in vacuum. In such a case, the direction of the shift corresponds to the sign of $\Delta m_{31}^2$, as well as to the CP phase of the neutrino/antineutrino beam.
\begin{figure}[!t]
\begin{center}
\includegraphics[width=\textwidth]{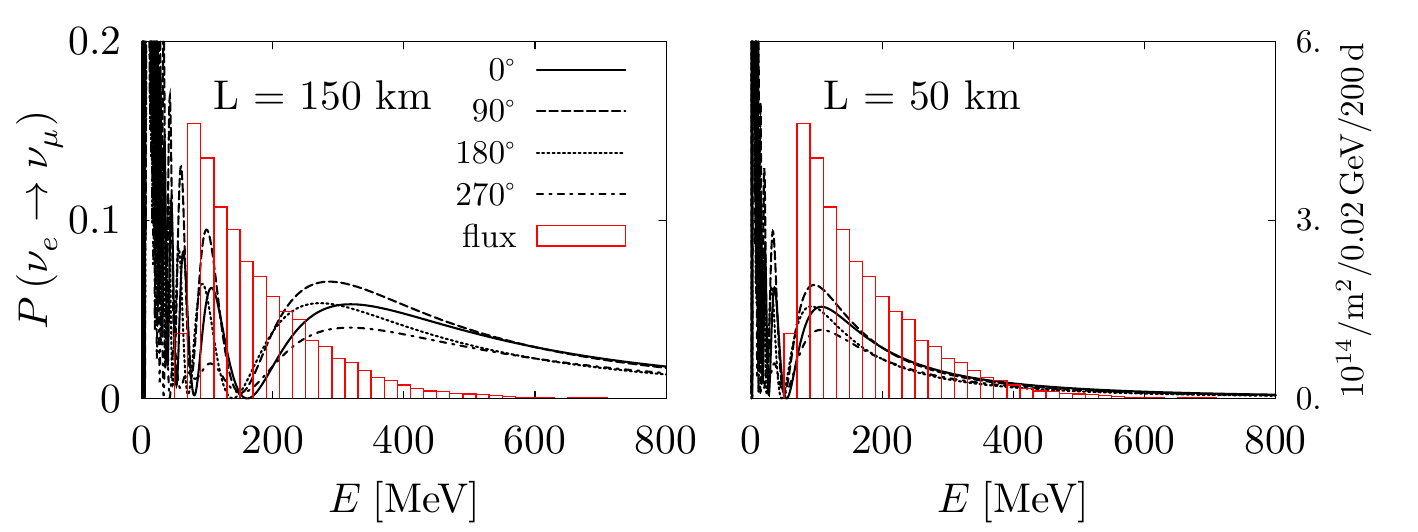}
\end{center}
\caption{\label{probability} The transition probability for the $\nu_e \rightarrow \nu_\mu$ channel, presented for the values $\delta_\text{CP} =$ 0$^\circ$, 90$^\circ$, 180$^\circ$ and 270$^\circ$. The probabilities are shown as functions of the neutrino energy $E =$ 0...800~MeV. In the left panel, the probabilities are shown for the experimental configurtion where the baseline length $L$ is optimized for the second oscillation maximum ($L =$ 150~km). In the right panel, the probabilities are correspondingly shown for the first oscillation maximum ($L =$ 50~km). For convenience, the simulated electron neutrino fluxes for MOMENT are shown with the red boxes. The fluxes are given in units of 10$^{14}$/m$^2$/\,0.02~GeV/\,200~d.}
\end{figure}

In this work, we focus on analysing the physics reach of MOMENT. The simulation of MOMENT is done with the General Long-Baseline Simulator (GLoBES) library~\cite{Huber:2004ka,Huber:2007ji} following the simulation details that were defined in Ref.~\cite{Tang:2017qen}. The relevant oscillation probabilities for MOMENT are presented in Fig.~\ref{probability}, where the oscillation probabilities corresponding to $\bar{\nu}_\mu \rightarrow \bar{\nu}_e$ and $\nu_e \rightarrow \nu_\mu$ transitions are plotted for the true values $\delta_\text{CP} =$ 0$^\circ$, 90$^\circ$, 180$^\circ$ and 270$^\circ$ with solid, dashed, dotted and dot-dashed lines, respectively. The electron neutrino fluxes used in our simulation of MOMENT are illustrated with the solid red lines. The fluxes are given in units of $\nu$/m$^2$/0.02~GeV/200~d and defined at the source.

As is apparent from Fig.~\ref{probability}, the neutrino fluxes used in our simulation of MOMENT can be sensitive to either the first or second oscillation maximum. The electron neutrino fluxes shown in the figure peak at $E \simeq$ 100~MeV, corresponding to the second oscillation maximum at the default baseline of $L =$ 150~km. In principle, the experimental setup could be optimized for the first oscillation maximum by moving the detector closer to the beam facility, such that $L =$ 50~km. This configuration could alternatively be sought by configuring the neutrino beam to peak at the energies $E \simeq$ 300~MeV.

Fig.~\ref{probability} illustrates the effect of varying the $\delta_\text{CP}$ parameter in the appearance probability $P (\nu_e \rightarrow \nu_\mu)$. In the left panel of the figure, the probability for $\delta_\text{CP} =$ 0$^\circ$ is nearly three times larger than the probability for $\delta_\text{CP} =$ 270$^\circ$, when neutrino energies near the second oscillation maximum are considered. In the right panel, on the other hand, the variation near the first oscillation maximum is less significant, though the wider shape of the oscillation maximum can be expected to yield a greater number of events.

\section{Simulation details}
\label{Section:3}

In this section we describe the experimental setup of MOMENT that is assumed for the simulations. The schematic design of the beam facility is presented in Ref.~\cite{Cao:2014bea}.

MOMENT entails a novel idea of producing a very intense neutrino-antineutrino beam at relatively low energies. In the plan, MOMENT includes a 1.5~GeV proton linac and a 10~mA proton driver, and it is aimed to produce a 15~MW proton beam in a continuous wave. The proton beam is proposed to hit a mercury jet-target, creating a beam of charged mesons mostly consisting of pions and kaons. The number of protons expected to hit the jet-target is 1.1$\times$10$^{24}$ per year. Super-conducting solenoids are then used to select pions of the desired energy range to be led to a decay tunnel. The selected pions decay in the tunnel into positively and negatively charged muons, which are further led to a 600~m decay tunnel where they can decay into neutrinos and antineutrinos. Whereas the positively charged muons decay into positrons and $\nu_e \, \bar{\nu}_\mu$ pairs, the negatively charged muons decay into electrons and $\bar{\nu}_e \, \nu_\mu$ pairs. The resulting beam of neutrinos and antineutrinos is expected to peak roughly at 200~MeV energy.

In the original proposal~\cite{Cao:2014bea}, the beam facility is proposed to be placed at the China Spallation Neutron Source whereas the location of the far detector is suggested to be near the JUNO far detector site, constituting a baseline of 150~km length. A far detector of approximately 500~kton mass has been suggested for this purpose. In previous works (see e.g. Ref.~\cite{Tang:2017qen}), a Water \v{C}erenkov detector with Gd-doping has been considered due to its excellent performance at low energies.

MOMENT benefits from well-constrained backgrounds from the source thanks to its muon-decay-based production mechanism. The main sources of background in MOMENT are charge mis-identification, detector imperfections, atmospheric neutrinos and neutral current backgrounds~\cite{Blennow:2015cmn,Tang:2017qen}. All of these background components set limitations to the experiments physics reach, and thus require a careful study.

In this work, we simulate a 10-year-long run in MOMENT, shared evenly between the $\mu^+$ run and the $\mu^-$ run. The simulation details concerning the 500-kton Water \v{C}erenkov detector with Gd-doping as well as the beam facility are summarized in Table~\ref{Experiment parameters}. The information of the detector design follows the description given in Ref.~\cite{Agostino:2012fd}, Gd-doping technology in Ref.~\cite{Beacom:2003nk}, and atmospheric neutrino background suppression in Refs.~\cite{Campagne:2006yx} and~\cite{FernandezMartinez:2009hb}. The neutrino and antineutrino fluxes are based on a Monte Carlo simulation provided by the accelerator working group~\cite{Tang:2017qen}, assuming 1.1$\times$10$^{24}$~POT per year (corresponding to 3.8$\times$10$^{21}$ neutrinos per year). The cross sections for the charged current and neutral current processes are taken from Ref.~\cite{Paschos:2001np}, respectively. In order to make a comparison or a combination of different experiments,  we also include the simulations of T2K and NO$\nu$A as well as of DUNE, T2HK and T2HKK. We take the simulation files for T2K~\cite{Huber:2002mx,Itow:2001ee,Ishitsuka:2005qi} and NO$\nu$A~\cite{Ambats:2004js,Yang_2004} from the GLoBES website~\cite{GLoBES}, whilst the files for DUNE are obtained from Ref.~\cite{Alion:2016uaj} and those of T2HK and T2HKK from the respective collaboration. The details of each long baseline experiment are summarized in Table~\ref{Experiment parameters} and Appendix~\ref{A1}.

\begin{table}
\caption{\label{Experiment parameters}The experimental inputs assumed in the simulation of MOMENT and the long baseline experiments T2K, NO$\nu$A, DUNE, T2HK and T2HKK. The energy resolution is given in terms of $\sigma_e$, which corresponds to the width of the Gaussian distribution.}
\begin{center}
\resizebox{\textwidth}{!}{%
\begin{tabular}{ccccccc}
\hline\hline
\rule{0pt}{3ex}Experiment & T2K & NO$\nu$A & DUNE & T2HK & MOMENT & T2HKK \\ \hline
\rule{0pt}{3ex}Location & Japan & USA & USA & Japan & China & Japan \\ 
\rule{0pt}{3ex}Status & operating & operating & under construction & proposed & proposed & proposed \\ 
\rule{0pt}{3ex}Accelerator facility & J-PARC & Fermilab & Fermilab & J-PARC & MOMENT & J-PARC \\ 
\rule{0pt}{3ex}Beam power & 770~kW & 700~kW & 1.07~MW & 1.3~MW & 15~MW & 1.3~MW \\ 
\rule{0pt}{3ex}Expected POT & 7.8$\times$10$^{21}$ & 3.6$\times$10$^{21}$ & 1.47$\times$10$^{21}$ & 1.56$\times$10$^{22}$ & 1.1$\times$10$^{25}$ & 1.56$\times$10$^{22}$ \\ 
\rule{0pt}{3ex}Baseline length & 295~km & 810~km & 1300~km & 295~km & 150~km & 295~km \& 1100~km\\ 
\rule{0pt}{3ex}Off-axis angle & 2.5$^\circ$ & 0.8$^\circ$ & 0$^\circ$ & 2.5$^\circ$ & 0$^\circ$ & 2.5$^\circ$ \& 1.5$^\circ$\\ 
\rule{0pt}{3ex}Detector technology & W.C. & L.Sc. & L.Ar. & W.C. \& W.C. & W.C. & W.C. \& W.C. \\ 
\rule{0pt}{3ex}Fiducial mass & 22.5~kt & 14~kt & 40~kt & 187~kt \& 187~kt & 500~kt & 187~kt \& 187~kt \\ 
\rule{0pt}{3ex}Running times (years) & 2+6 & 3+3 & 5+5 & 2.5+7.5 & 5+5 & 2.5+7.5 \\ 
\rule{0pt}{3ex}Energy window (GeV) & [0.4,~1.2] & [1.0,~3.5] & [0.5,~18.0] & [0.1,~10.0] & [0.1,~0.8] & [0.1,~10.0]\\
\rule{0pt}{3ex}Energy bins & 20 & 25 & 71 & 12 & 20 & 12 \\
\rule{0pt}{3ex}$\sigma_e = \alpha \, E + \beta \, \sqrt{E} + \gamma$ & $\gamma =$ 0.085~GeV & \parbox[c]{3cm}{$\beta =$ 10\% ($e$-like)\\$\beta =$ 5\% ($\mu$-like)} & migration matrices & migration matrices & $\alpha =$ 8.5\% & migration matrices \\
\rule{0pt}{3ex}References & Ref.~\cite{Abe:2011ks} & Ref.~\cite{Ayres:2007tu} & Ref.~\cite{Acciarri:2015uup} & Ref.~\cite{Abe:2015zbg} & Ref.~\cite{Cao:2014bea} & Ref.~\cite{Abe:2016ero} \\
\hline\hline
\end{tabular}}
\end{center}
\end{table}

Regarding the simulation of the 500~kt Gd-doped WC detector, we adopt the same methodology as what was used in Ref.~\cite{Tang:2017qen}. Throughout our work, we assume a 40\% reconstruction efficiency for electron-like events and 50\% efficiency for muon-like events, respectively. As far as the background is concerned, we assign a 0.3\% acceptance rate for wrong-charge particles being tagged as signal events and 0.25\% rate for events originating from neutral current interactions. 

For the atmospheric neutrino background, we assume the same treatment as was done in Ref.~\cite{Agostino:2012fd}. We calculate the number of events from the atmospheric neutrino background in each energy bin as follows:
\begin{equation}
N_i = T \, N_\text{nucl} \, f \int_{E_\text{min}}^{E_\text{max}} dE \int_{{E'}_\text{min}}^{{E'}_\text{max}} dE' \, \phi(E) \, \sigma(E) \, R(E,E'),
\label{sim:01}
\end{equation}
where $i$ is the number of energy bin, $T$ is the total runtime of the experiment and $N_\text{nucl}$ is the number of target nucleons in the detector. The true and reconstructed energies are denoted as $E$ and $E'$, respectively. Whereas $E_\text{min} =$ 100~MeV and $E_\text{max} =$ 800~MeV are the boundaries of the considered energy range, $E^{'}_\text{min}$ and $E^{'}_\text{max}$ denote the boundaries of the energy bin. The atmospheric neutrino fluxes are given by $\phi(E)$, while $\sigma(E)$ stands for the cross sections and $R(E,E')$ is the energy resolution function. The atmospheric neutrino flux is constrained by the suppression factor $f$, which we take to be 2.2$\times$10$^{-3}$ after angular cuts and measuring the atmospheric background during the beam-off time.\footnote{In Refs.~\cite{Blennow:2015cmn,FernandezMartinez:2009hb}, the suppression factor is given by $N\,l/L$, where $N$ is the number of ion or proton bunches, $L$ is the length of the decay volume, while $l$ gives the distance between two bunches. In both definitions, smaller suppression factors correspond to better reduction of the atmospheric neutrino background.}

From now on, we briefly describe how the $\chi^2$ values are calculated and what parameter values are used in the simulation. The central values for the oscillation parameters are taken from the corresponding best-fit values, which are summarized in Table~\ref{bounds}.
\begin{table}[!t]
\caption{\label{bounds} The current best-fit values and 1$\,\sigma$ uncertainties for the neutrino oscillation parameters~\cite{NuFit:4-1}. These values are shown for both mass orderings and are taken from a recent global analysis. Note that $\Delta m_{3l}^2$ stands for $\Delta m_{31}^2$ in normal ordering (NO) and $\Delta m_{32}^2$ in inverted ordering (IO).}
\begin{center}
\begin{tabular}{ccccc}\hline\hline
Parameter & Central value $\pm$ 1\,$\sigma$ (NO) \,\,\, & Central value $\pm$ 1\,$\sigma$ (IO) \\ \hline
\rule{0pt}{3ex}$\theta_{12}$ ($^\circ$) & 33.82 $\pm$ 0.78 & 33.82 $\pm$ 0.78 \\ 
\rule{0pt}{3ex}$\theta_{13}$ ($^\circ$) & 8.60 $\pm$ 0.13 & 8.64 $\pm$ 0.13 \\ 
\rule{0pt}{3ex}$\theta_{23}$ ($^\circ$) & 48.6 $\pm$ 1.4 & 48.8 $\pm$ 1.2 \\ 
\rule{0pt}{3ex}$\delta_\text{CP}$ ($^\circ$) & 221 $\pm$ 39 & 282 $\pm$ 25 \\ 
\rule{0pt}{3ex}$\Delta m_{21}^2$ (10$^{-5}$ eV$^2$) & 7.39 $\pm$ 0.21 & 7.39 $\pm$ 0.21 \\ 
\rule{0pt}{3ex}$\Delta m_{3l}^2$ (10$^{-3}$ eV$^2$) & 2.528 $\pm$ 0.031 & -2.510 $\pm$ 0.031 \\ \hline\hline
\end{tabular}
\end{center}
\end{table}

The statistical part of the $\chi^2$ function is calculated with the well-known Poissonian~\cite{Huber:2004ka}:
\begin{equation}
\chi^2 (\vec{\omega}, \vec{\omega}_0, \zeta_\text{sg}, \zeta_\text{bg}) = \sum_{i=1}^{20} 2 \left[ T_i - O_i \left( 1 + \ln \frac{T_i}{O_i} \right) \right],
\label{eq:01}
\end{equation}
where $\vec{\omega} \equiv$ ($\theta_{12}$, $\theta_{13}$, $\theta_{23}$, $\delta_\text{CP}$, $\Delta m_{21}^2$, $\Delta m_{31}^2$) and $\vec{\omega_0}$ are the so called test and true values of the oscillation parameters, $O_i$ and $T_i$ are the event numbers corresponding to $\vec{\omega}$ and $\vec{\omega_0}$, and $\zeta_\text{sg}$ and $\zeta_\text{bg}$ are the nuisance parameters describing the overall normalization error for the signal and background events, respectively. $O_i$ and $T_i$ are obtained from $\vec{\omega}$ and $\vec{\omega_0}$ as follows: $O_i (\vec{\omega}_0) = N_i^\text{sg} (\vec{\omega}_0) + N_i^\text{bg} (\vec{\omega}_0)$ and $T_i (\vec{\omega}, \zeta_\text{sg}, \zeta_\text{bg}) = (1 + \zeta_\text{sg}) \, N_i^\text{sg} (\vec{\omega}) + (1 + \zeta_\text{bg}) \, N_i^\text{bg} (\vec{\omega})$, where $N_i^\text{sg}$ and $N_i^\text{bg}$ represent the expected numbers for signal and background events, respectively.

The $\chi^2$ value corresponding to a neutrino signal, e.g. from events originating from $\nu_e \rightarrow \nu_\mu$ oscillation, is estimated from the Poissonian shown in Eq.~(\ref{eq:01}) after including the systematics with the conventional pull method and applying the experimental uncertainties with the priors:
\begin{equation}
\chi^2 (\vec{\omega}_0) = \min_{ \vec{\omega}, \zeta_\text{sg}, \zeta_\text{bg} } \left[ \chi^2_\text{stat}\left( \vec{\omega}, \vec{\omega}_0, \zeta_\text{sg}, \zeta_\text{bg} \right) + \frac{\zeta_\text{sg}^2}{\sigma_\text{sg}^2} + \frac{\zeta_\text{bg}^2}{\sigma_\text{bg}^2} + \sum_{k=1}^6 \left( \frac{\omega_k - (\omega_0)_k}{\sigma_k} \right)^2 \right],
\label{eq:06}
\end{equation}
where $\sigma_\text{sg}$ and $\sigma_\text{sg}$ are the errors associated with the nuisance parameters $\zeta_\text{sg}$ and $\zeta_\text{bg}$, respectively. The minimization in Eq.~(\ref{eq:06}) is first performed over the nuisance parameters $\zeta_\text{sg}$ and $\zeta_\text{bg}$ and then over the test values $\vec{\omega}$.

The overall $\chi^2$ distributions are calculated by summing the $\chi^2$ values corresponding to the relevant oscillation channels in the simulated experiment. Each channel is assigned with a signal and background error $\sigma_\text{sg}$ and $\sigma_\text{bg}$, whereas the priors are taken from Table~\ref{bounds}. In our simulation of MOMENT, we set $\sigma_\text{sg} =$ 0.025 and $\sigma_\text{bg} =$ 0.05 for electron-like events, and $\sigma_\text{sg} =$ 0.05 and $\sigma_\text{bg} =$ 0.05 for muon-like events.

The precision on the $\delta_\text{CP}$ parameter can be conveniently expressed with $\Delta \chi^2 = \chi^2 - \chi^2_\text{min}$, which is defined as the difference of $\chi^2$ calculated at an arbitrary test value of $\delta_\text{CP}$ and at a value that is taken to be its true value. Throughout this work, we calculate $\Delta \chi^2$ by minimizing over all oscillation parameters except for $\delta_\text{CP}$, and assuming the neutrino mass ordering and $\theta_{23}$ octant be known. 

We express our results in terms of $\Delta \delta_\text{CP}$, which we define as the average of the higher and lower deviations from the expected central value, that is,
\begin{equation}
\Delta \delta_\text{CP} =  \frac{\delta_\text{CP}^\text{upper} - \delta_\text{CP}^\text{lower}}{2},
\label{eq:07}
\end{equation}
where $\delta_\text{CP}^\text{upper}$ and $\delta_\text{CP}^\text{lower}$ are the upper and lower bounds, respectively, that can be expected to be attainable in the simulated neutrino experiment. In this work, we focus on the 1$\,\sigma$ confidence level, at which it is sufficient to approximate $\Delta \chi^2$ as a $\chi^2$ distribution of one degree of freedom. Here $\delta_\text{CP}^\text{upper}$ and $\delta_\text{CP}^\text{lower}$ are determined as the fit values that satisfy the requirement $\Delta \chi^2 =$ 1.

\section{Precision measurement of $\delta_\text{CP}$ in MOMENT}
\label{Section:4}

We present the results of our work in this section. In Section~\ref{Section:4.1}, we show sensitivity to the Dirac CP phase $\delta_\text{CP}$ in MOMENT. We compare the sensitivities of MOMENT with those of currently running long baseline oscillation experiments T2K and NO$\nu$A, and proposed next-generation experiments T2HK and DUNE. In Section~\ref{Section:4.2}, we investigate the impact of several experimental parameters on the $\delta_\text{CP}$ sensitivity in MOMENT. More specifically, we study the effects of statistics, systematics and beam polarity. We also discuss the effect of the second oscillation maximum in the measurement of $\delta_\text{CP}$. Finally, in Section~\ref{Section:4.3}, we study the complementarity of MOMENT and the long baseline neutrino experiments in the precision of $\delta_\text{CP}$.

\subsection{Sensitivity to the $\delta_\text{CP}$ parameter in MOMENT}
\label{Section:4.1}

\begin{figure}[!t]
\begin{center}
\includegraphics[width=\textwidth]{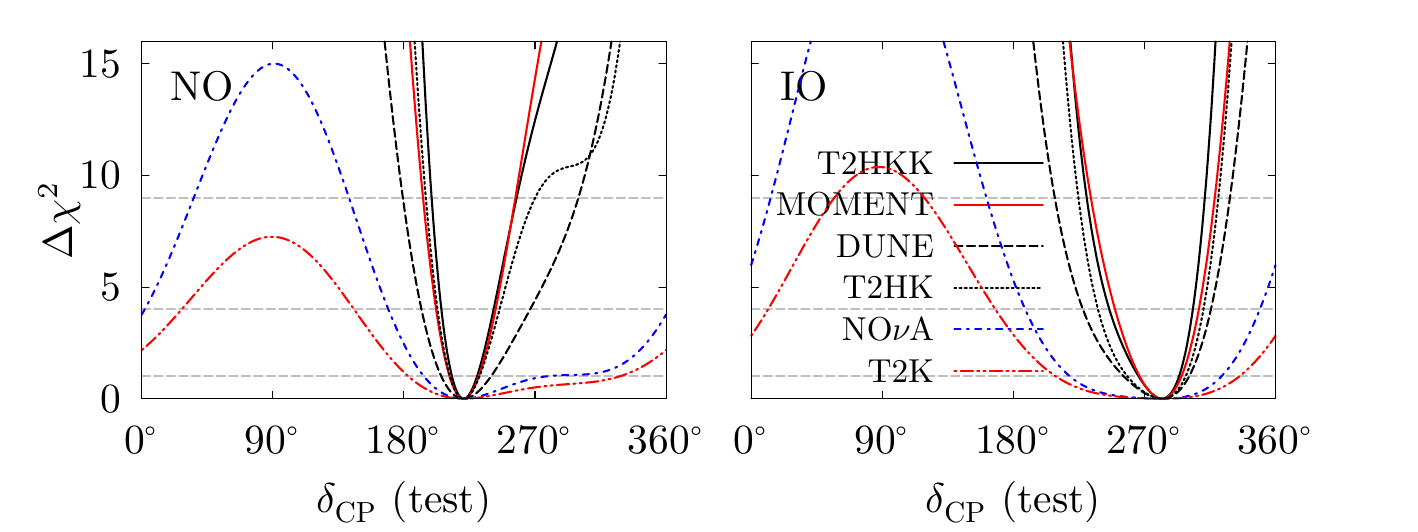}
\end{center}
\caption{\label{ChiSquare} The sensitivity to the $\delta_\text{CP}$ parameter in neutrino oscillation experiments. The $\Delta \chi^2$ distributions of the $\delta_\text{CP}$ parameter are shown for its allowed values $\delta_\text{CP} =$ 0$^\circ$...360$^\circ$. The 1$\,\sigma$, 2$\,\sigma$ and 3$\,\sigma$ confidence levels are shown with the dashed lines corresponding to $\Delta \chi^2 =$ 1, 4 and 9, respectively. The sensitivities are shown for MOMENT, T2HK, DUNE, NO$\nu$A and T2K, and for both normal (NO) and inverted mass orderings (IO).}
\end{figure}

The sensitivity to the $\delta_\text{CP}$ parameter measurement is illustrated in Fig.~\ref{ChiSquare}. The figure presents the values of $\Delta \chi^2$ as function of $\delta_\text{CP} =$ 0$^\circ$...360$^\circ$, which are inserted as test values in Eq.~(\ref{eq:06}). The true value of $\delta_\text{CP}$ is taken to be $\delta_\text{CP} =$ 221$^\circ$ when the normal ordering (NO, $m_1 < m_2 \ll m_3$) is assumed, and $\delta_\text{CP} =$ 282$^\circ$ when the ordering is taken to be inverted\footnote[1]{The values of $\delta_\text{CP} =$ 221$^\circ$ (NO) and $\delta_\text{CP} =$ 282$^\circ$ (IO) correspond to the current best-fit values from fitting to the global neutrino oscillation data. The fits include the atmospheric neutrino data from Super-Kamiokande~\cite{NuFit:4-1,Esteban:2018azc}.} (IO, $m_3 \ll m_1 < m_2$). The $\Delta \chi^2$ distributions are shown for $\delta_\text{CP}$ in T2K, NO$\nu$A, T2HK, DUNE, MOMENT and T2HKK. The results for T2K and NO$\nu$A as well as for the future experiments are prediction forecasts. For convenience, the 1$\,\sigma$, 2$\,\sigma$ and 3$\,\sigma$ confidence level contours are shown with the lines $\Delta \chi ^2 =$ 1, 4 and 9, respectively.

Our results predict excellent sensitivities to the precision of the $\delta_\text{CP}$ parameter in MOMENT. As can be seen in Fig.~\ref{ChiSquare}, the sensitivities of MOMENT can be expected to be comparable with those of future long baseline experiments DUNE and T2HK, as well as the T2HKK alternative, and superior to those of currently running long baseline experiments NO$\nu$A and T2K.

\subsection{Optimizing the $\delta_\text{CP}$ measurement in MOMENT}
\label{Section:4.2}

Though the physics reach of MOMENT has been studied in a number of references~\cite{Blennow:2015cmn,Bakhti:2016prn,Tang:2017qen,Tang:2018rer}, it should be stressed that the experimental setup of MOMENT is still open. In this section we investigate the relevance of four different characteristics related to the experimental setup in the precision measurement of the $\delta_\text{CP}$ parameter. More specifically, we study how statistics, systematics, beam sharing and baseline length affect $\Delta \delta_\text{CP}$ in MOMENT. We also investigate how much the results depend on the assumptions that were made on the systematic uncertainties, detector efficiencies and the size of the background.

\subsubsection{Statistical uncertainties}
\label{Section:4.2.1}
\begin{figure}[!t]
\begin{center}
\includegraphics[width=0.5\textwidth]{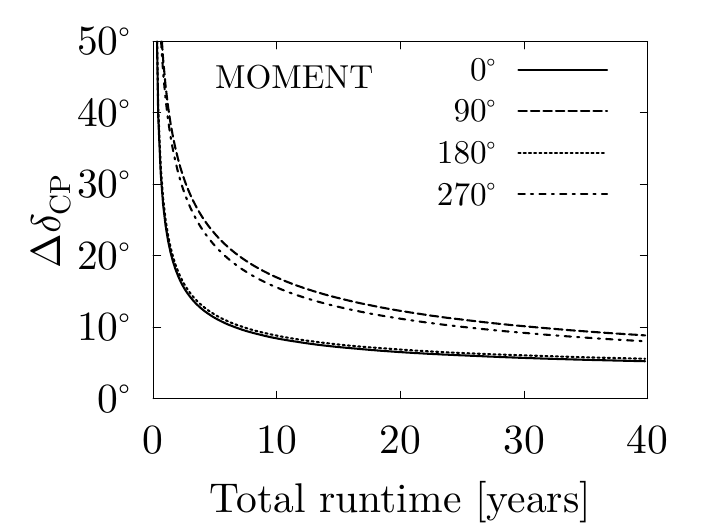}
\end{center}
\caption{\label{RunningTime} The effect of statistics on the precision measurement of $\delta_\text{CP}$. The 1$\,\sigma$ confidence level uncertainties for $\delta_\text{CP}$ are shown as a function of the total running time of the experiment. The $\Delta \delta_\text{CP}$ distributions are shown for the true values $\delta_\text{CP}=$ 0, 90$^\circ$, 180$^\circ$ and 270$^\circ$. The results are independent of the mass ordering.}
\end{figure}

We first want to understand the relevance of statistics on the determination of $\Delta \delta_\text{CP}$ in MOMENT. We investigate this by studying the dependence of $\Delta \delta_\text{CP}$ on the running time of the experiment. In Fig.~\ref{RunningTime}, $\Delta \delta_\text{CP}$ is shown as function of the total running time of the experiment, shared evenly between the $\mu^+$ and $\mu^-$ modes. Since $\Delta \delta_\text{CP}$ also depends on the true value of $\delta_\text{CP}$, we present the $\Delta \delta_\text{CP}$ contours for $\delta_\text{CP} =$ 0$^\circ$, 90$^\circ$, 180$^\circ$ and 270$^\circ$. We remark that NO and IO yield nearly identical numerical results.

Fig.~\ref{RunningTime} shows how increasing the total running time of the experiment, where the muon beam is shared evenly between the $\mu^+$ and $\mu^-$ modes, leads to an improved result on $\Delta \delta_\text{CP}$. For the CP conserving values $\delta_\text{CP} =$ 0$^\circ$ or 180$^\circ$, the sensitivity becomes rather difficult to improve after 10 years of running. For example, doubling the running time from 10 years to 20 years improves the estimate on $\Delta \delta_\text{CP}$ by less than 2$^\circ$. Increasing the running time further after 20 years leads to smaller improvements. For the maximally CP violating values $\delta_\text{CP} =$ 90$^\circ$ and 270$^\circ$, on the other hand, similar increases in running time leads to more significant improvements. All results are obtained at 1$\,\sigma$ CL.

One should keep in mind that increasing the running time is but one method to increase the statistics of the experiment. In principle, doubling the total running time is equivalent to doubling the fiducial mass of the detector or the intensity of the neutrino beam.

\subsubsection{Systematic uncertainties}
\label{Section:4.2.2}

\begin{figure}[!t]
\begin{center}
\includegraphics[width=\textwidth]{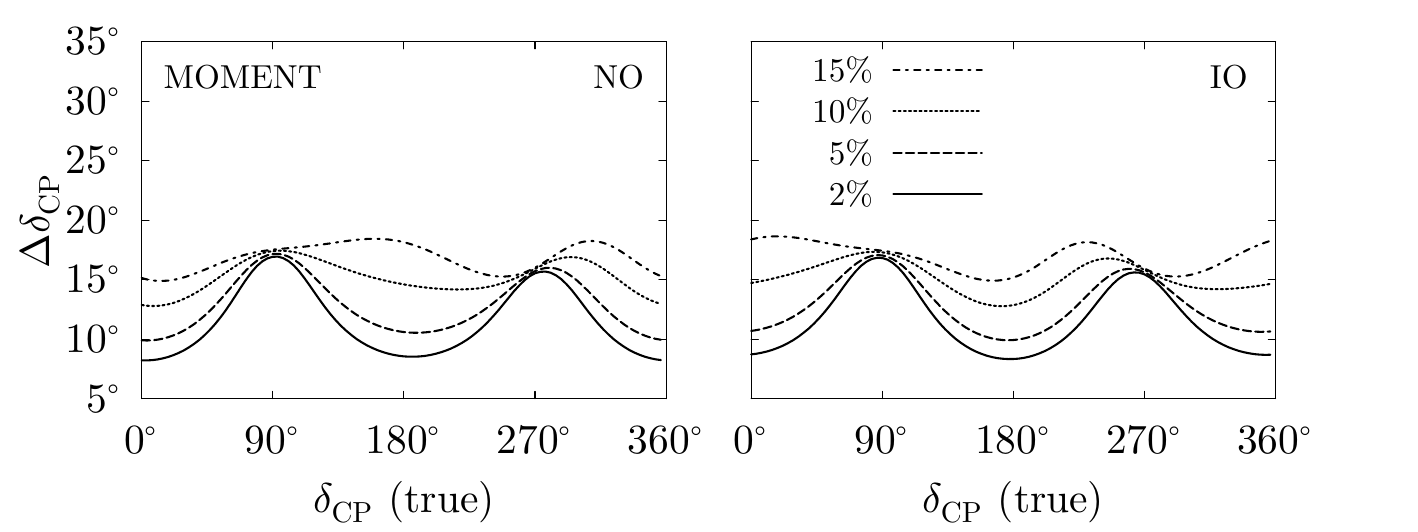}
\end{center}
\caption{\label{Systematics} The effect of the systematics on the precision measurement of $\delta_\text{CP}$. The 1$\,\sigma$ confidence level uncertainties for $\delta_\text{CP}$ are shown as a function of values $\delta_\text{CP} =$ 0$^\circ$...360$^\circ$. The uncertainties are shown for four different scenarios where the systematic error on the signal normalization is set to either $\sigma_{\zeta_\text{sg}}=$ 2\%, 5\%, 10\% or 15\%. The results are shown for both normal (NO) and inverted ordering (IO).}
\end{figure}

We next want to study the impact of the systematics on the precision of $\delta_\text{CP}$ in MOMENT. The importance of systematics is illustrated in Fig.~\ref{Systematics}, where $\Delta \delta_\text{CP}$ is plotted at 1$\,\sigma$ CL as a function of $\delta_\text{CP}$. The results are presented for four different values of $\sigma_{\zeta_\text{sg}}$, which we choose to be 2\%, 5\%, 10\% and 15\%. These errors correspond to the systematics arising from the overall normalization of the signal events, and it is set for relevant oscillation channels considered in this work. We assume a uniform signal normalization error in all signal events.

As is apparent from Fig.~\ref{Systematics}, the estimate on $\Delta \delta_\text{CP}$ at 1$\,\sigma$ CL correlates with the considered systematics significantly for all values except the maximally CP violating values $\delta_\text{CP} =$ 90$^\circ$ and 270$^\circ$. This behaviour can be tracked back to the origin of CP sensitivity in $\sin \delta_\text{CP}$ (see Section~\ref{Section:2}). Hence, one may say that whilst $\delta_\text{CP}$ precision is dominated by systematics near the CP conserving values 0$^\circ$ and 180$^\circ$, any improvements on the precision must be obtained through the increase of statistics near maximally CP violating values 90$^\circ$ and 270$^\circ$, respectively.

We note that the simulation setup we defined for MOMENT in Section~\ref{Section:3} leads to a sensitivity that is very similar to the 2\% curve in Fig.~\ref{Systematics}. The default setup assumes a 2.5\% systematic error on all appearance channels and 5\% error on disappearance channels. The overlapping with the 2\% error curve therefore highlights the importance of the appearance channels in the precision measurement on $\delta_\text{CP}$.

We investigated the dependence of the results on the assumptions made on the systematic uncertainties, detector efficiencies and size of the background. The results are presented in Appendix~\ref{Appendix B}. We find that any changes to the detector efficiencies or background shift the $\Delta \delta_\text{CP}$ distributions either to the better or the worse. The shift is the most significant for $\delta_\text{CP} =$ 90$^\circ$ and 270$^\circ$, and the least significant for $\delta_\text{CP} =$ 0$^\circ$ and 180$^\circ$, respectively.

\subsubsection{Beam polarity}
\label{Section:4.2.3}

\begin{figure}[!t]
\begin{center}
\includegraphics[width=0.5\textwidth]{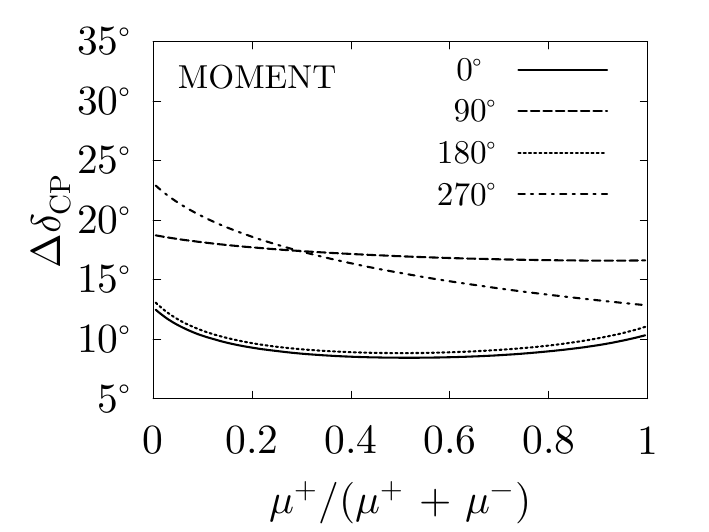}
\end{center}
\caption{\label{BeamPolarity} The effect of beam polarity on the precision measurement of $\delta_\text{CP}$. The 1$\,\sigma$ confidence level uncertainties for $\delta_\text{CP}$ are shown as a function of the fraction at which MOMENT operates in $\mu^+$ mode, whereas the rest of the operation is run in $\mu^-$ mode. The uncertainties are shown for $\delta_\text{CP} =$ 0$^\circ$, 90$^\circ$, 180$^\circ$ and 270$^\circ$. The results are independent of the mass ordering.}
\end{figure}

We next set our focus on the importance of the beam polarity in the precision of the $\delta_\text{CP}$ parameter. MOMENT is known to produce neutrinos by generating beams of $\mu^+$ and $\mu^-$, which are assumed to be produced in equal amounts in our baseline setup. This is one of the characteristics that can be changed in MOMENT. In Fig.~\ref{BeamPolarity}, we show how $\Delta \delta_\text{CP}$ (1$\,\sigma$ CL) changes for $\delta_\text{CP} =$ 0$^\circ$, 90$^\circ$, 180$^\circ$ and 270$^\circ$ when the beam time is shared between the two polarities in asymmetrical running times. For this purpose, we define the quantity $\mu^+ / (\mu^+ + \mu^-)$ as the fraction at which MOMENT runs in $\mu^+$ mode, whereas the rest of the 10-year-long running time is dedicated to $\mu^-$ mode.

The following observations can be made from Fig.~\ref{BeamPolarity}: For the CP conserving values $\delta_\text{CP} =$ 0$^\circ$ and 180$^\circ$, we find a mild preference for the symmetrical running of 5+5 years. When $\delta_\text{CP} =$ 90$^\circ$ or 270$^\circ$, a pure $\mu^+$ beam run is preferred instead, though it should be noted that the preference for the 10+0--year running time is not significant in the case of $\delta_\text{CP} =$ 90$^\circ$.

We searched for the reason why the $\Delta \delta_\text{CP}$ distributions take different shapes for $\delta_\text{CP} =$ 0$^\circ$, 90$^\circ$, 180$^\circ$ and 270$^\circ$ in Fig.~\ref{BeamPolarity}. We find that the sensitivity to $\Delta \delta_\text{CP}$ is dominated by specific oscillation channels for different values of $\mu^+ / (\mu^+ + \mu^-)$. Whereas events from the $\nu_e \rightarrow \nu_\mu$ appearance channel constitutes the sensitivity when $\mu^+ / (\mu^+ + \mu^-)$ is close to one, the sensitivity to $\delta_\text{CP}$ arises mainly from $\nu_\mu \rightarrow \nu_e$ events when the value of $\mu^+ / (\mu^+ + \mu^-)$ is close to zero. The impact of the corresponding antineutrino channels is negligible due to low signal-to-background ratio. For the true values $\delta_\text{CP} =$ 0$^\circ$ and 180$^\circ$, the effect of $\nu_e \rightarrow \nu_\mu$ and $\nu_\mu \rightarrow \nu_e$ channels is nearly symmetric, yielding nearly equal contributions to the $\Delta \delta_\text{CP}$ distribution. For the true values of $\delta_\text{CP} =$ 90$^\circ$ and 270$^\circ$, however, there exists a subtle asymmetry in favour of $\nu_e \rightarrow \nu_\mu$ events, giving $\Delta \delta_\text{CP}$ lower values for longer running times in the positive muon mode. While the preference is mild for $\delta_\text{CP} =$ 90$^\circ$, the effect is noticeable when $\delta_\text{CP} =$ 270$^\circ$.

We also investigated how changing the systematic uncertainties in the signal and background events as well as the detector coefficients and the size of the background changes the 1$\,\sigma$ confidence level uncertainties in Fig.~\ref{BeamPolarity}. We made the following remarks:
\begin{itemize}
  \item Increasing the signal error $\sigma_\text{sg}$ up to 15\% leads to negligible changes in $\Delta \delta_\text{CP}$ when $\delta_\text{CP} =$ 90$^\circ$ or 270$^\circ$. For $\delta_\text{CP} =$ 0$^\circ$ and 180$^\circ$, however, $\Delta \delta_\text{CP}$ increased significantly.
  \item When the background error $\sigma_\text{bg}$ was increased up to 15\%, the change in $\Delta \delta_\text{CP}$ was less than 1$^\circ$ for all considered values of $\delta_\text{CP}$.
  \item Changing the detector efficiencies lead to a shift in $\Delta \delta_\text{CP}$ distributions that has milder effect when the $\mu^+ / (\mu^+ + \mu^-)$ ratio is close to one. Conversely, the change in $\Delta \delta_\text{CP}$ is stronger when the ratio is close to zero. We observed similar effects for altering the size of the background.
  \item All of the changes in $\Delta \delta_\text{CP}$ conserved the shape mainly due to fact that the variations of experimental parameters were always imposed symmetrically on all of the oscillation channels. Were any of the neutrino appearance channels favoured over the others, the change might have also shown in the shapes.
\end{itemize}

We remark that the results presented in Fig.~\ref{BeamPolarity} were obtained when NO was assumed. We find the difference to be negligible when IO is considered.

\subsubsection{First and second oscillation maxima}
\label{Section:4.2.4}

We also study the effect of the first and second oscillation maxima in the precision on $\delta_\text{CP}$ parameter. The first oscillation maximum occurs when the condition $\Delta \, L / 2 \equiv L\,\Delta m_{31}^2 / 4\,E = \pi / 2$ is met. For the proposed baseline length of MOMENT, $L \simeq$ 150~km, the second oscillation maximum occurs for neutrino energies $E \simeq$ 300~MeV. The second oscillation maximum follows similarly from the equation $\Delta \, L / 2 = 3\pi / 2$, which indicates $E \simeq$ 100~MeV. We can therefore expect MOMENT to be sensitive to both the first and second oscillation maxima, with more events coming from the second maximum.

\begin{figure}[!t]
\begin{center}
\includegraphics[width=0.5\textwidth]{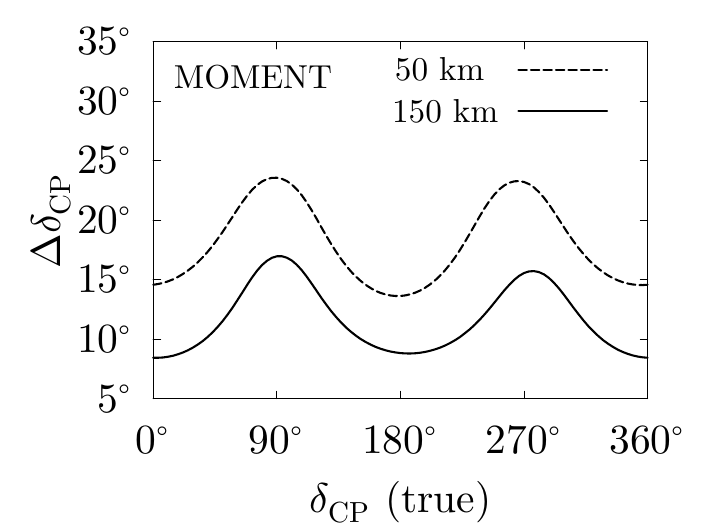}
\end{center}
\caption{\label{SecondMaximum} The effect of the first oscillation maximum on the precision measurement of $\delta_\text{CP}$. The 1$\,\sigma$ confidence level uncertainty of $\delta_\text{CP}$ is shown for two baseline lengths, $L =$ 50~km and 150~km, which correspond to the experimental configurations where MOMENT is optimized for the first and second oscillation maximum, respectively. The results are independent of the mass ordering.}
\end{figure}

In theory, the experimental setup of MOMENT can be optimized to match the first oscillation maximum instead of the second one. In our simulation of MOMENT, the mean value of the neutrino energies is $E \simeq$ 200~MeV. For these energies, the first oscillation maximum occurs at $L \simeq$ 50~km and the second maximum at $L \simeq$ 150~km. We therefore choose to compare the sensitivity to $\Delta \delta_\text{CP}$ at 1$\,\sigma$ CL in MOMENT at baseline lengths 50~km and 150~km. The results are shown for NO in Fig.~\ref{SecondMaximum}. We find the results nearly identical when IO is considered.
 
In Fig.~\ref{SecondMaximum}, the $\Delta \delta_\mathrm{CP}$ distributions are shown in two scenarios. Whereas the solid line illustrates the $\Delta \delta_\mathrm{CP}$ contour (1$\,\sigma$ CL) as function of $\delta_\text{CP}$ values at 150~km baseline length, the dashed curve corresponds to the $\Delta \delta_\mathrm{CP}$ contour at 50~km length with the same integrated luminosity. Whilst the first case corresponds to the baseline setup of MOMENT, the latter case corresponds to the situation where MOMENT is optimized for the first oscillation maximum. It is evident from the figure that the second oscillation maximum leads to a greater sensitivity on $\Delta \delta_\text{CP}$ for all available values of $\delta_\text{CP}$. This is true despite the fact that at 50~km, the expected number of neutrino and antineutrino events at the far detector is expected to be nine times higher in comparison to the 150~km baseline length.

\subsection{Complementarity of MOMENT and superbeam experiments in the measurement of $\delta_\text{CP}$}
\label{Section:4.3}

In previous sections, we have investigated how well the parameter $\delta_\text{CP}$ can be measured in MOMENT, and how the sensitivities of MOMENT compare with those of the currently running and planned long baseline neutrino experiments. In this part of the work, we study how precisely the value of $\delta_\text{CP}$ can be determined after MOMENT finishes running.

The complementarity of the superbeam experiments and MOMENT play an important role in the determination of the $\delta_\text{CP}$ value at a high 1$\,\sigma$ CL precision. Although MOMENT has access to a greater number of neutrino oscillation channels and lower beam backgrounds than superbeam experiments, the greater statistics provided by DUNE and T2HK near the first oscillation maximum improve the expected resolution on $\delta_\text{CP}$ throughout its currently allowed values. T2HKK can be expected to improve the precision further, as it provides events from both the first and second oscillation maximum~\cite{Abe:2016ero}. The inclusion of the currently running experiments T2K and NO$\nu$A, however, can be expected to yield a non-significant contribution due to their relatively limited statistics.

\begin{figure}[!t]
\begin{center}
\includegraphics[width=\textwidth]{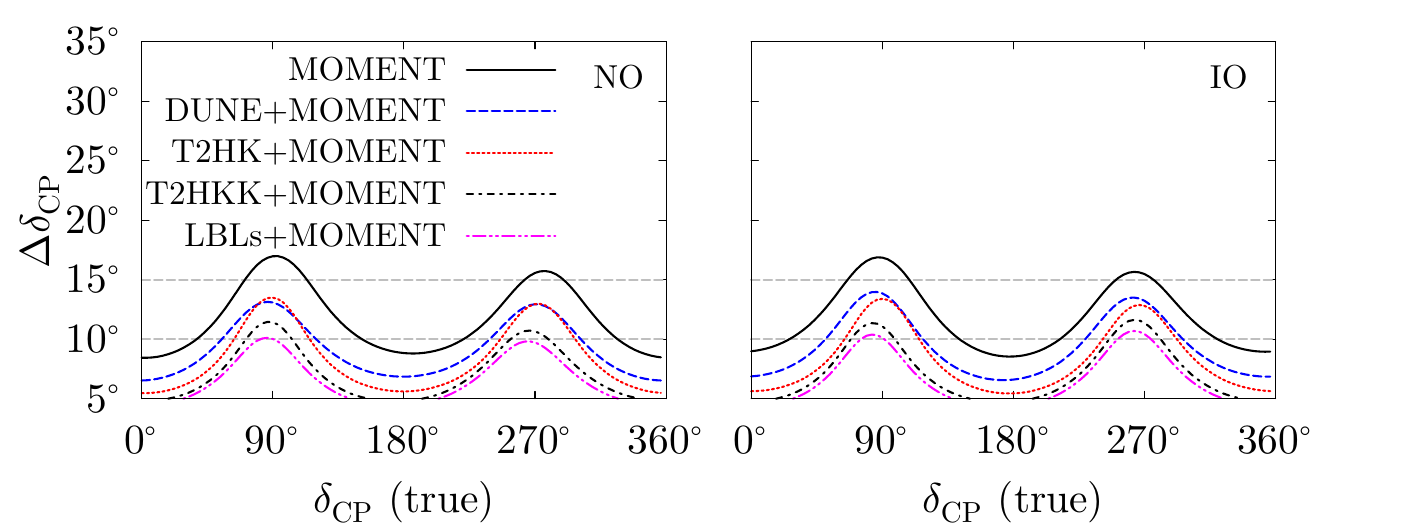}
\end{center}
\caption{\label{CombinedData} The 1$\,\sigma$ confidence level uncertainty of $\delta_\text{CP}$ shown for four different combinations of neutrino oscillation data. The results are shown for the currently allowed values of $\delta_\text{CP}$ in normal ordering (NO) and inverted ordering (IO). The grey dashed lines represent the benchmark values of $\Delta \delta_\text{CP} =$ 10$^\circ$ and 15$^\circ$. Here LBLs+MOMENT corresponds to the combined run of the DUNE, MOMENT and T2HKK experiments.}
\end{figure}

In Fig.~\ref{CombinedData}, we show the 1$\,\sigma$ CL uncertainty of $\delta_\text{CP}$ as function of $\delta_\text{CP}$ values when the simulated data from MOMENT is analyzed jointly with that of either DUNE (dashed line), T2HK (dotted) or T2HKK (dot-dashed), and in the event when the data from MOMENT, DUNE and T2HKK are analyzed together (dot-dot-dashed). We also present the expected sensitivity to $\Delta \delta_\text{CP}$ at 1$\,\sigma$ CL also for the event when only the data from MOMENT is analyzed. T2K and NO$\nu$A are not included in the simulation. We have ascertained that their contribution is negligible.

Basing on the results shown in Fig.~\ref{CombinedData}, we find that MOMENT has an appreciable sensitivity to measure the $\delta_\text{CP}$ parameter at 1$\,\sigma$ CL. After 10 years of taking data, MOMENT can be expected to determine the value of $\delta_\text{CP}$ by greater than 20$^\circ$ resolution throughout the $\delta_\text{CP}$ value spectrum, and by 15$^\circ$ or better for 79\% of the currently allowed values, regardless of the mass ordering. At the 10$^\circ$ discrimination level, we find MOMENT to be able to reach the desired resolution for 38\% and 36\% of the values in NO and IO, respectively.

The 1$\,\sigma$ CL constraints on the true value of $\delta_\text{CP}$ can be improved further when the data from MOMENT is analyzed jointly with those of the currently planned long baseline experiments T2HK and DUNE. When the data from DUNE and MOMENT are analyzed together, we find the experiments to be able to measure the value of $\delta_\text{CP}$ by 10$^\circ$ or better at 1$\,\sigma$ CL for 63\% regardless of the mass ordering. When the data from DUNE is replaced with that of T2HK, on the other hand, the 10$^\circ$ coverage extends to 69\% for both NO and IO. Replacing the data from T2HK with that of T2HKK instead improves the result to 82\% in NO and 80\% in IO, respectively. The most precise result can be expected from the scenario where DUNE, MOMENT and T2HKK are able to conduct their planned runs, in which case the value of $\delta_\text{CP}$ can be determined at 1$\,\sigma$ CL by at least 10$^\circ$ for 98\% and 90\% of the considered $\delta_\text{CP}$ values for NO and IO, respectively.

In our concluding remark, we briefly comment on the prospects of measuring the value of $\delta_\text{CP}$ in the next-generation accelerator experiments when the true value of $\delta_\text{CP}$ is 270$^\circ$. When the neutrino masses are normally ordered, we find MOMENT to be able to determine the value of $\delta_\text{CP}$ by 15.6$^\circ$ at 1$\,\sigma$ CL, whereas the inclusion of either the DUNE or T2HK data could improve the precision to 12.9$^\circ$, and of T2HKK to 10.7$^\circ$, respectively. Combining the data from MOMENT, DUNE and T2HKK could yield an overall resolution of 10.1$^\circ$, and 9.3$^\circ$ when MOMENT is let to run entirely in $\mu^+$ mode. All results are obtained at 1$\,\sigma$ CL. We obtained similar results for the inverted mass ordering.

\section{Summary}
\label{Section:6}

We have investigated the physics reach of MOMENT in the measurement of Dirac CP phase $\delta_\text{CP}$, and compared its sensitivities to those of the planned superbeam experiments T2HK, T2HKK and DUNE, as well as the running T2K and NO$\nu$A. With the proposed setup~\cite{Cao:2014bea}, we find the sensitivities in MOMENT to be comparable with those of T2HK and T2HKK and greater than the sensitivities simulated in DUNE, T2K and NO$\nu$A. 

We have studied the effect of several experimental characteristics on the precision measurement of $\delta_\text{CP}$, in order to identify potential bottlenecks and opportunities for improvement in MOMENT. We have investigated the effect of statistical and systematic uncertainties, changing the fraction of the positive/negative beam polarity in the total running time, and adjusting the baseline length to match the first or second oscillation maximum on the precision measurement of $\delta_\text{CP}$. Our results are summarized in Table~\ref{resultstable}.

The current design of MOMENT provides the advantage of delivering a low-background beam where the systematic uncertainties can be assumed to be relatively low. This in turn gives the experiment a leverage on measuring the value of $\delta_\text{CP}$ near the CP conserving values of $\delta_\text{CP} =$ 0$^\circ$ and 180$^\circ$. An exception to this feature is seen near the maximally CP violating values of $\delta_\text{CP} =$ 90$^\circ$ and 270$^\circ$, where systematics only play a small role and improvements to the CP precision must be achieved by increasing the statistics.

The CP precision in MOMENT has been previously studied in Refs.~\cite{Blennow:2015cmn} and \cite{Bakhti:2016prn}, where the simulated neutrino and antineutrino fluxes peak between 100...200~MeV. In the present work, the results are presented for a configuration where the beam is optimized for the second oscillation maximum. The beam fluxes described in this paper are also used in Refs.~\cite{Tang:2017qen} and \cite{Tang:2018rer}. We find that this setup leads to a greater resolution on $\delta_\text{CP}$.

\begin{table}[!t]
\begin{center}
\caption{\label{resultstable}  The impact of statistical error, systematic uncertainties, changing the fraction of the positive/negative beam polarity in the total running time and adjusting the beam energy and baseline setup to match the first and second oscillation maximum on $\Delta \delta_\text{CP}$ for $\delta_\text{CP} =$ 0$^\circ$, 90$^\circ$, 180$^\circ$ and 270$^\circ$ at 1$\,\sigma$ confidence level. The values are shown for the normal mass ordering. The final column shows references to the relevant figures.}
\begin{tabular}{c c c c c c}\hline\hline
$\delta_\text{CP}$ & 0$^\circ$ & 90$^\circ$ & 180$^\circ$ & 270$^\circ$ & Figure \\ \hline
\rule{0pt}{3ex} Running times & & & & & Fig.~\ref{RunningTime} \\
\rule{0pt}{3ex} 10 years & 8.4$^\circ$ & 17.0$^\circ$ & 8.2$^\circ$ & 15.6$^\circ$\\
\rule{0pt}{3ex} 40 years & 5.2$^\circ$ & 8.9$^\circ$ & 5.6$^\circ$ & 8.0$^\circ$\\ \hline
\rule{0pt}{3ex} Systematics & & & & & Fig.~\ref{Systematics} \\
\rule{0pt}{3ex} 2\% & 8.2$^\circ$ & 16.9$^\circ$ & 8.6$^\circ$ & 15.5$^\circ$\\
\rule{0pt}{3ex} 15\% & 15.1$^\circ$ & 17.5$^\circ$ & 18.1$^\circ$ & 16.0$^\circ$\\ \hline
\rule{0pt}{3ex} Beam polarity & & & & & Fig.~\ref{BeamPolarity} \\
\rule{0pt}{3ex} 50\%/50\% & 8.4$^\circ$ & 17.0$^\circ$ & 8.8$^\circ$ & 15.6$^\circ$\\
\rule{0pt}{3ex} 100\%/0\% & 10.3$^\circ$ & 16.6$^\circ$ & 11.0$^\circ$ & 12.8$^\circ$\\ \hline
\rule{0pt}{3ex} Oscillation maximum & & & & & Fig.~\ref{SecondMaximum} \\
\rule{0pt}{3ex} 1$^\text{st}$ max. & 14.6$^\circ$ & 23.6$^\circ$ & 13.6$^\circ$ & 23.2$^\circ$\\
\rule{0pt}{3ex} 2$^\text{nd}$ max. & 8.4$^\circ$ & 17.0$^\circ$ & 8.8$^\circ$ & 15.6$^\circ$\\ \hline\hline
\end{tabular}
\end{center}
\end{table}

Altogether, we find that MOMENT has an excellent potential to constrain the $\delta_\text{CP}$ parameter value to about 10$^\circ$ or better at 1$\,\sigma$ confidence level, when its data is analyzed together with those of the proposed superbeam experiments T2HKK and DUNE. The same resolution can also be achieved in MOMENT alone by increasing its statistics by a factor of four.

We conclude this work with the following remarks:
\begin{itemize}
	\item Our simulation shows excellent precision in the measurement of $\delta_\text{CP}$ in MOMENT. With the proposed setup, the sensitivities in MOMENT are comparable with those of T2HKK and T2HK and greater than those DUNE.
	\item For the current best-fit values, MOMENT can reach the approximate precision of 10.0$^\circ$ at 1$\,\sigma$ confidence level in normal ordering and 14.3$^\circ$ in inverted ordering, respectively.
	\item Many of the experimental parameters in MOMENT are still open for discussion. We have found that in the case of $\delta_\text{CP} =$ 270$^\circ$, the highest precision on $\delta_\text{CP}$ can be obtained when the experiment runs entirely in the $\mu^+$ mode.
	\item MOMENT can improve the $\delta_\text{CP}$ bounds from those of T2HK and DUNE significantly. When the three are analysed together, we find that MOMENT can improve $\delta_\text{CP}$ precision to 11.8$^\circ$ or better at 1$\,\sigma$ confidence level, regardless of the mass ordering or the true value of $\delta_\text{CP}$. When T2HKK is considered instead of T2HK, the precision improves to 10.7$^\circ$, respectively.
\end{itemize}

\acknowledgments
This work is supported in part by the National Natural Science Foundation of China under Grant Nos. 11505301 and 11881240247, and by the university support based on National SuperComputer Center-Guangzhou (74130-31143408). We would like to thank the accelerator working group of MOMENT for providing the neutrino flux files. We also thank Dr. Nick W. Prouse for the T2HKK simulation package.
 
\section{Appendix}
\appendix
\section{The long baseline neutrino experiments considered in this work}
\label{A1}

In this appendix, we list the long baseline experiments that have been considered in this work. We also describe how the simulation of each experiment was created.

\subsection{T2K}

The Tokai-to-Kamioka (T2K)~\cite{Abe:2011ks} experiment is a currently running long-baseline neutrino oscillation experiment in Japan with a 295~km baseline. The neutrino and antineutrino beams of predominantly muon flavour are generated via pion decay in the J-PARC facility in Tokai, where a 30~GeV proton beam is directed at a graphite target. The neutrino beam and subsequently the antineutrino beam are first studied at the near detectors ND280 and INGRID, which are both located approximately 280 meters from the target. The far detector facility in Kamioka is Super-Kamiokande, which is a Water \v{C}erenkov detector of 22.5~kton of fiducial mass located at 2.5$^\circ$ off-axis from the beam. The produced neutrino and antineutrino beams peak at roughly 600~MeV, which corresponds to the first oscillation maximum for $\nu_\mu \rightarrow \nu_e$ and $\bar{\nu}_\mu \rightarrow \bar{\nu}_e$ channels. T2K began operating in 2011, and it is expected to yield a total of 7.8$\times$10$^{21}$ protons-on-target (POT) with 2 years of neutrino run and 6 years of antineutrino run.

In our simulation of T2K, we assume a full 2+6--year running time in neutrino and antineutrino modes, 22.5~kton of fiducial mass and 0.77~MW beam power. The studied energy range is 0.4~GeV...1.2~GeV, which is divided into 20 energy bins of equal size. The energy resolution is taken to be a constant 0.085~GeV to account the Fermi motion. For more details regarding the simulation, see the GLoBES website~\cite{GLoBES}.

\subsection{NO$\nu$A}

The NuMI Off-axis $\nu_e$ Appearance (NO$\nu$A)~\cite{Ayres:2007tu} is another on-going long baseline neutrino oscillation experiment, which is currently operating in the United States. NO$\nu$A consists of the NuMI beam facility and a near detector in Fermilab, and of a 14~kt Liquid Scintillator detector located in Ash River, Minnesota. The far detector is located 810~km from the source at 14.6~mrad off-axis, where the first oscillation maximum occurs for $\nu_\mu \rightarrow \nu_e$ and $\bar{\nu}_\mu \rightarrow \bar{\nu}_e$ approximately at 2~GeV. The near detector on the other hand, consists of 200~ton and it is located 1~km from the source along the axis. Both the near and far detectors are highly active tracking calorimeters segmented with hundreds of PVP cells. NO$\nu$A began running in 2014 and it is expected to reach the exposure of 3.6$\times$10$^{21}$~POT after running 3 years in neutrino mode and 3 years in antineutrino mode.

In our simulation of NO$\nu$A, we adopt the simulation files that are available on the GLoBES website~\cite{GLoBES}. We changed the fiducial mass of the detector and the expected POT number to 14~kton and 3.6$\times$10$^{21}$, respectively.

\subsection{T2HK and T2HKK}

The Tokai-to-Hyper-Kamiokande (T2HK) experiment~\cite{Abe:2015zbg} is a next-generation neutrino oscillation experiment and a successor to T2K. T2HK will use a beam setup similar to its predecessor T2K, that is, a 2.5 degree off-axis beam from J-PARC. Before T2HK starts operating, the beam facility will be upgraded to 1.3~MW power and Hyper-Kamiokande~\cite{Hyper-Kamiokande:2016dsw} will take over Super-Kamiokande as the far detector. The new detector will consist of two Water \v{C}erenkov tanks, equivalent to 187~kton fiducial mass each. The experiment is expected to run 2.5 years in neutrino mode and 7.5 years in antineutrino mode, expecting a total exposure of 1.56$\times$10$^{22}$~POT. An alternative of placing the second detector in Korea instead has also been considered under the T2HKK proposal~\cite{Abe:2016ero}.

In this work, we conduct the simulation for T2HK and T2HKK using simulation files that were provided by the T2HK collaboration. In the case of T2HK, we assume a staged approach where the experiment runs the first six years with one detector and the other four years with two detectors. Both detector vessels are taken to be located at 295~km from the source. For the simulation of T2HKK, on the other hand, we assume a full 10-year run with the second detector vessel placed in Korea at 1100~km from the source, whilst the first detector vessel is located in Kamioka at 295~km. For the Korea detector, we assume a 1.5$^\circ$ off-axis angle.

\subsection{DUNE}

The Deep Underground Neutrino Experiment (DUNE)~\cite{Acciarri:2016crz,Acciarri:2015uup,Strait:2016mof,Acciarri:2016ooe} is a next generation long-baseline neutrino experiment currently under construction in the United States. Its design features the LBNF neutrino beam in Fermilab and a 40~kton far detector located in the Sanford Underground Research Facility at the Homestake mine, located 1300~km from the source. The detector is based on the Liquid Argon Time Projection Chamber technology. The beam consists of muon neutrinos and antineutrinos peaking roughly at 2.5~GeV energy. A total exposure of 1.47$\times$10$^{21}$ POT is expected for DUNE after running 3.5~years in neutrino mode and 3.5~years in antineutrino mode. A multipurpose near detector is located approximately at 500~m from the source. DUNE is expected to begin its run in 2026.

In our simulation of DUNE, we adopt the simulation files that were provided by the DUNE collaboration, see Ref.~\cite{Alion:2016uaj} for more details.

\section{The effects of detector efficiencies and atmospheric neutrino background on the results}
\label{Appendix B}

In this appendix, we provide more details on how different assumptions on the systematic uncertainties, detector efficiencies and background size affect the $\Delta \delta_\text{CP}$ distributions we obtained for MOMENT. This discussion is necessary, as many of the technical details of MOMENT are still uncertain. Among the unsettled questions are at least the performance of the Gd-doped Water \v{C}erenkov technology and the suppression of the atmospheric neutrino background. 
\begin{figure}[!h]
\begin{center}
\begin{subfigure}{0.72\textwidth}
\includegraphics[width=\linewidth]{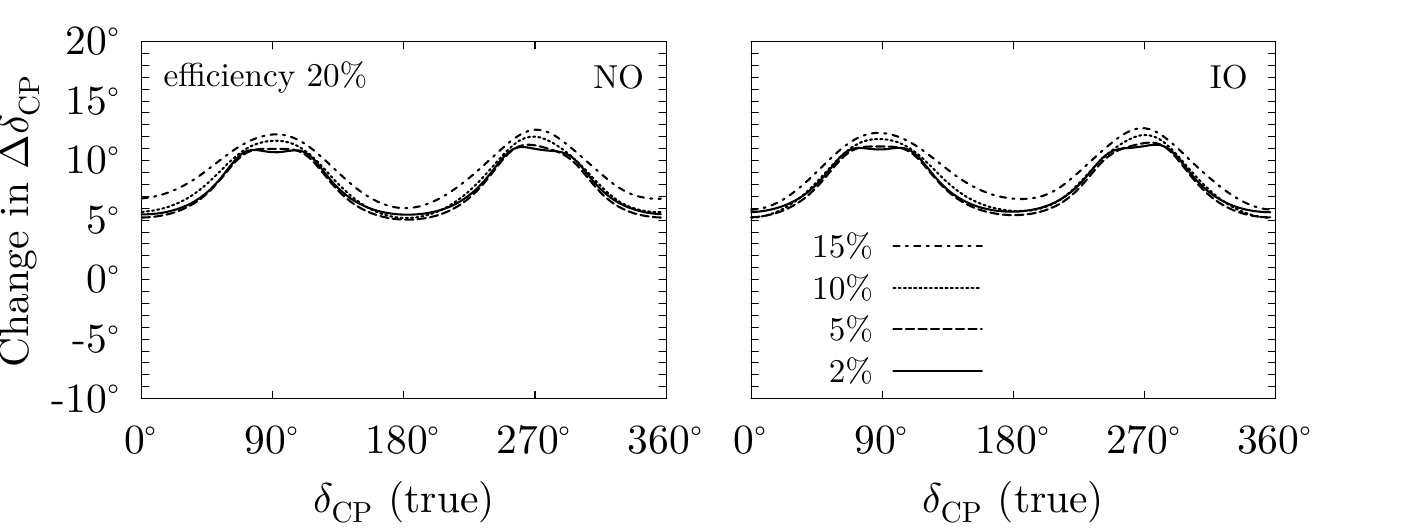}
\caption{Detector efficiencies 20\%} \label{Syst1}
\end{subfigure}

\medskip
\begin{subfigure}{0.72\textwidth}
\includegraphics[width=\linewidth]{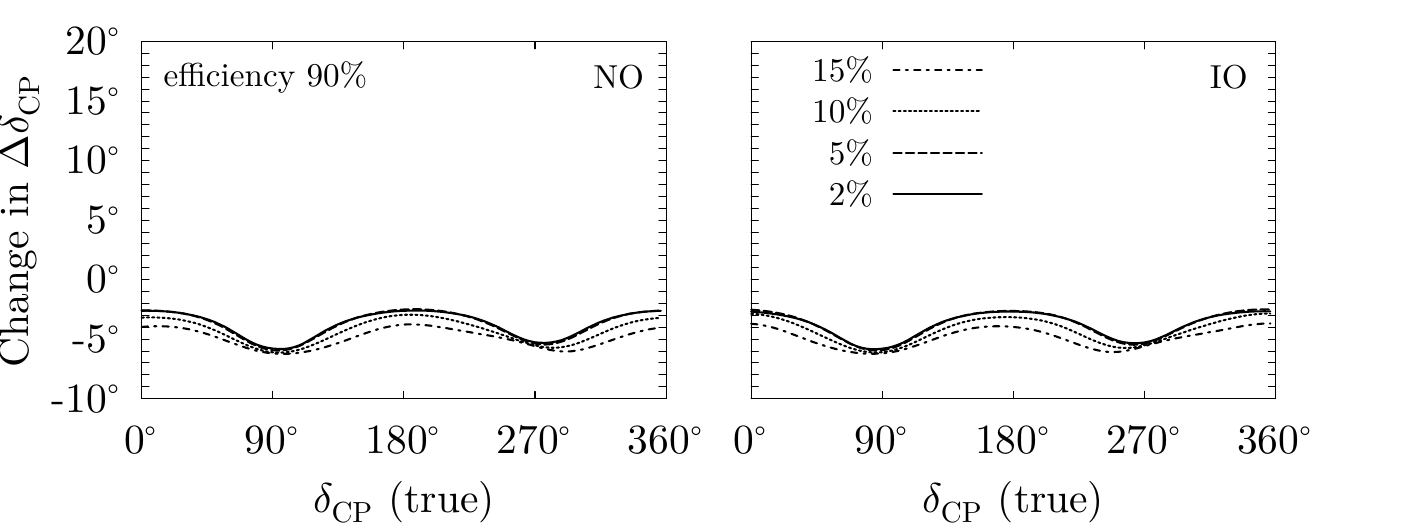}
\caption{Detector efficiencies 90\%} \label{Syst2}
\end{subfigure}

\medskip
\begin{subfigure}{0.72\textwidth}
\includegraphics[width=\linewidth]{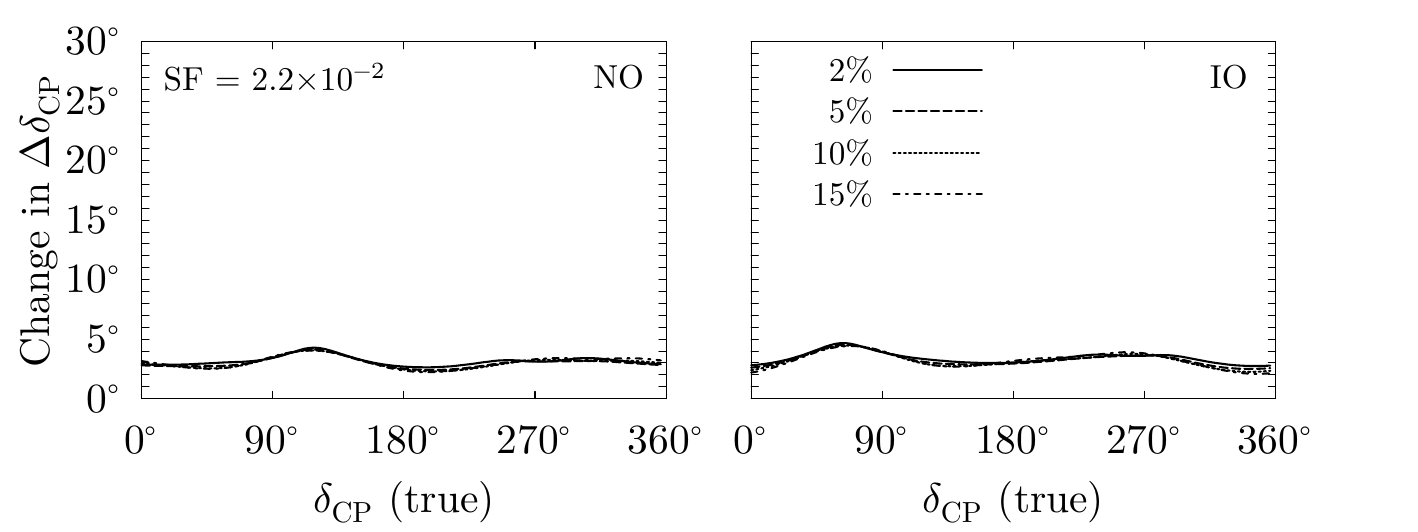}
\caption{Suppression factor 2.2$\times$10$^{-2}$} \label{Syst3}
\end{subfigure}

\medskip
\begin{subfigure}{0.72\textwidth}
\includegraphics[width=\linewidth]{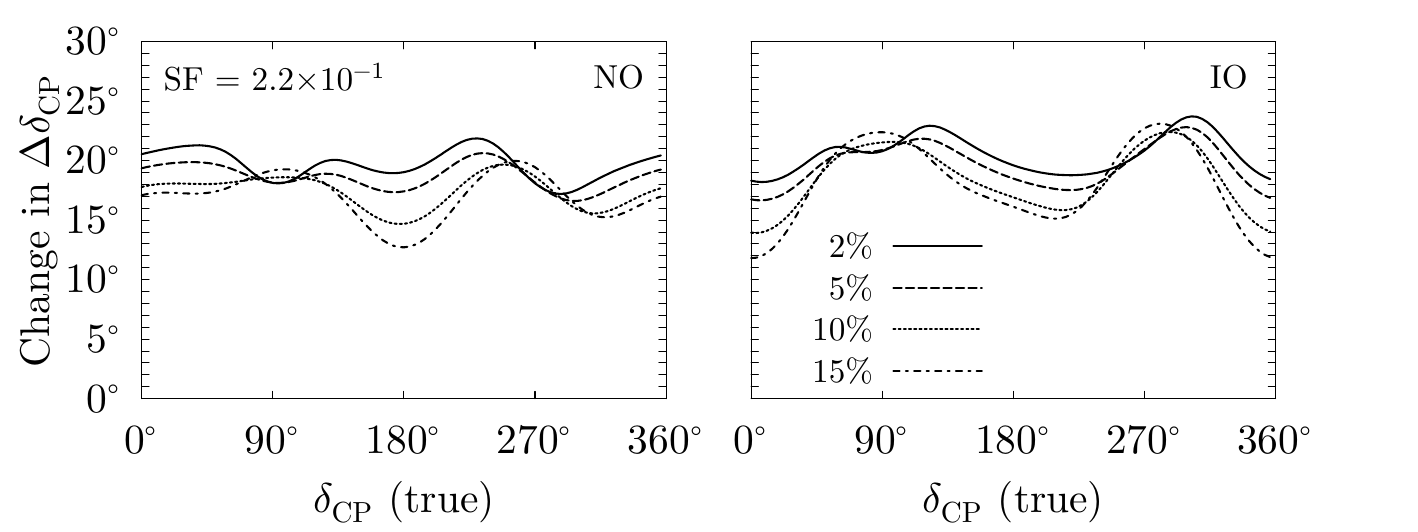}
\caption{Suppression factor 2.2$\times$10$^{-1}$} \label{Syst4}
\end{subfigure}

\caption{The effect of detector efficiencies and atmospheric neutrino background on $\Delta \delta_\text{CP}$ in MOMENT when the signal errors are set to 2\%, 5\%, 10\% or 15\%. Deviations from the results obtained with the baseline setup are presented for $\Delta \delta_\text{CP}$ when detector efficiencies are set to (a) 20\% or (b) 90\%, and when the atmospheric neutrino background is suppressed by a factor of (c) 2.2$\times$10$^{-2}$ and (d) 2.2$\times$10$^{-2}$ instead of 2.2$\times$10$^{-3}$, respectively. The atmospheric neutrino background is calculated with Eq.~(\ref{sim:01}). The results are shown for both normal ordering (NO) and inverted ordering (IO). \label{fig:A2}}
\end{center}
\end{figure}

In Fig.~\ref{fig:A2}, we present the deviations that occur in $\Delta \delta_\text{CP}$ distributions when the systematic uncertainties of the experiment are varied under different assumptions on the detector efficiencies and the atmospheric neutrino background. In panels~\ref{Syst1} and~\ref{Syst2}, the change in $\Delta \delta_\text{CP}$ is calculated when the detector efficiencies in signal channels are set to 20\% and 90\%, respectively. Panels~\ref{Syst3} and~\ref{Syst4} on the other hand represent the situations where the suppression factor for the atmospheric neutrino background is taken to be 2.2$\times$10$^{-2}$ and 2.2$\times$10$^{-1}$. Both results are compared to that of the baseline setup, where the suppression factor is assumed to be 2.2$\times$10$^{-3}$. The atmospheric neutrino background and the suppression factor are defined in Eq.~(\ref{sim:01}) in Section~\ref{Section:3}.
\clearpage

\bibliographystyle{JHEP}
\bibliography{bibliography} 

\providecommand{\href}[2]{#2}\begingroup\raggedright\begin{thebibliography}{10}

\bibitem{Kajita:2000mr}
T.~Kajita and Y.~Totsuka, \emph{{Observation of atmospheric neutrinos}},
  \href{https://doi.org/10.1103/RevModPhys.73.85}{\emph{Rev. Mod. Phys.}
  {\bfseries 73} (2001) 85}.

\bibitem{Ahmad:2001an}
{\scshape SNO} collaboration, \emph{{Measurement of the rate of $\nu_e+d \to
  p+p+e^-$ interactions produced by $^8B$ solar neutrinos at the Sudbury
  Neutrino Observatory}},
  \href{https://doi.org/10.1103/PhysRevLett.87.071301}{\emph{Phys. Rev. Lett.}
  {\bfseries 87} (2001) 071301}
  [\href{https://arxiv.org/abs/nucl-ex/0106015}{{\ttfamily nucl-ex/0106015}}].

\bibitem{Pontecorvo:1957cp}
B.~Pontecorvo, \emph{{Mesonium and anti-mesonium}}, {\emph{Sov. Phys. JETP}
  {\bfseries 6} (1957) 429}.

\bibitem{Maki:1960ut}
Z.~Maki, M.~Nakagawa, Y.~Ohnuki and S.~Sakata, \emph{{A unified model for
  elementary particles}},
  \href{https://doi.org/10.1143/PTP.23.1174}{\emph{Prog. Theor. Phys.}
  {\bfseries 23} (1960) 1174}.

\bibitem{Kobayashi:1973fv}
M.~Kobayashi and T.~Maskawa, \emph{{CP Violation in the Renormalizable Theory
  of Weak Interaction}}, \href{https://doi.org/10.1143/PTP.49.652}{\emph{Prog.
  Theor. Phys.} {\bfseries 49} (1973) 652}.

\bibitem{NuFit:4-1}
``{NuFIT 4.1 (2019)}.'' \url{http://www.nu-fit.org/}.

\bibitem{Esteban:2018azc}
I.~Esteban, M.~C. Gonzalez-Garcia, A.~Hernandez-Cabezudo, M.~Maltoni and
  T.~Schwetz, \emph{{Global analysis of three-flavour neutrino oscillations:
  synergies and tensions in the determination of $\theta_{23}, \delta_{CP}$,
  and the mass ordering}},
  \href{https://doi.org/10.1007/JHEP01(2019)106}{\emph{JHEP} {\bfseries 01}
  (2019) 106} [\href{https://arxiv.org/abs/1811.05487}{{\ttfamily
  1811.05487}}].

\bibitem{deSalas:2018bym}
P.~F. De~Salas, S.~Gariazzo, O.~Mena, C.~A. Ternes and M.~Tortola,
  \emph{{Neutrino Mass Ordering from Oscillations and Beyond: 2018 Status and
  Future Prospects}},
  \href{https://doi.org/10.3389/fspas.2018.00036}{\emph{Front. Astron. Space
  Sci.} {\bfseries 5} (2018) 36}
  [\href{https://arxiv.org/abs/1806.11051}{{\ttfamily 1806.11051}}].

\bibitem{Capozzi:2018ubv}
F.~Capozzi, E.~Lisi, A.~Marrone and A.~Palazzo, \emph{{Current unknowns in the
  three neutrino framework}},
  \href{https://doi.org/10.1016/j.ppnp.2018.05.005}{\emph{Prog. Part. Nucl.
  Phys.} {\bfseries 102} (2018) 48}
  [\href{https://arxiv.org/abs/arXiv:1804.09678}{{\ttfamily
  arXiv:1804.09678}}].

\bibitem{King:2015aea}
S.~F. King, \emph{{Models of Neutrino Mass, Mixing and CP Violation}},
  \href{https://doi.org/10.1088/0954-3899/42/12/123001}{\emph{J. Phys.}
  {\bfseries G42} (2015) 123001}
  [\href{https://arxiv.org/abs/1510.02091}{{\ttfamily 1510.02091}}].

\bibitem{Petcov:2018snn}
S.~T. Petcov and A.~V. Titov, \emph{{Assessing the Viability of $A_4$, $S_4$
  and $A_5$ Flavour Symmetries for Description of Neutrino Mixing}},
  \href{https://doi.org/10.1103/PhysRevD.97.115045}{\emph{Phys. Rev.}
  {\bfseries D97} (2018) 115045}
  [\href{https://arxiv.org/abs/1804.00182}{{\ttfamily 1804.00182}}].

\bibitem{Ballett:2016daj}
P.~Ballett, S.~F. King, S.~Pascoli, N.~W. Prouse and T.~Wang,
  \emph{{Sensitivities and synergies of DUNE and T2HK}},
  \href{https://doi.org/10.1103/PhysRevD.96.033003}{\emph{Phys. Rev.}
  {\bfseries D96} (2017) 033003}
  [\href{https://arxiv.org/abs/1612.07275}{{\ttfamily 1612.07275}}].

\bibitem{Ghosh:2019sfi}
M.~Ghosh and T.~Ohlsson, \emph{{A comparative study between ESSnuSB and T2HK in
  determining the leptonic CP phase}},
  \href{https://arxiv.org/abs/1906.05779}{{\ttfamily 1906.05779}}.

\bibitem{Coloma:2012wq}
P.~Coloma, A.~Donini, E.~Fernandez-Martinez and P.~Hernandez, \emph{{Precision
  on leptonic mixing parameters at future neutrino oscillation experiments}},
  \href{https://doi.org/10.1007/JHEP06(2012)073}{\emph{JHEP} {\bfseries 06}
  (2012) 073} [\href{https://arxiv.org/abs/1203.5651}{{\ttfamily 1203.5651}}].

\bibitem{Winter:2003ye}
W.~Winter, \emph{{Understanding CP phase dependent measurements at neutrino
  superbeams in terms of bi-rate graphs}},
  \href{https://doi.org/10.1103/PhysRevD.70.033006}{\emph{Phys. Rev.}
  {\bfseries D70} (2004) 033006}
  [\href{https://arxiv.org/abs/hep-ph/0310307}{{\ttfamily hep-ph/0310307}}].

\bibitem{Smirnov:2018ywm}
M.~V. Smirnov, Z.~J. Hu, S.~J. Li and J.~J. Ling, \emph{{The possibility of
  leptonic CP-violation measurement with JUNO}},
  \href{https://doi.org/10.1016/j.nuclphysb.2018.05.003}{\emph{Nucl. Phys.}
  {\bfseries B931} (2018) 437}
  [\href{https://arxiv.org/abs/1802.03677}{{\ttfamily 1802.03677}}].

\bibitem{Cao:2014bea}
J.~Cao et~al., \emph{{Muon-decay medium-baseline neutrino beam facility}},
  \href{https://doi.org/10.1103/PhysRevSTAB.17.090101}{\emph{Phys. Rev. ST
  Accel. Beams} {\bfseries 17} (2014) 090101}
  [\href{https://arxiv.org/abs/1401.8125}{{\ttfamily 1401.8125}}].

\bibitem{Blennow:2015cmn}
M.~Blennow, P.~Coloma and E.~Fern\'{a}ndez-Martinez, \emph{{The MOMENT to
  search for CP violation}},
  \href{https://doi.org/10.1007/JHEP03(2016)197}{\emph{JHEP} {\bfseries 03}
  (2016) 197} [\href{https://arxiv.org/abs/1511.02859}{{\ttfamily
  1511.02859}}].

\bibitem{Bakhti:2016prn}
P.~Bakhti and Y.~Farzan, \emph{{CP-Violation and Non-Standard Interactions at
  the MOMENT}}, \href{https://doi.org/10.1007/JHEP07(2016)109}{\emph{JHEP}
  {\bfseries 07} (2016) 109}
  [\href{https://arxiv.org/abs/1602.07099}{{\ttfamily 1602.07099}}].

\bibitem{Tang:2017qen}
J.~Tang and Y.~Zhang, \emph{{Study of nonstandard charged-current interactions
  at the MOMENT experiment}},
  \href{https://doi.org/10.1103/PhysRevD.97.035018}{\emph{Phys. Rev.}
  {\bfseries D97} (2018) 035018}
  [\href{https://arxiv.org/abs/1705.09500}{{\ttfamily 1705.09500}}].

\bibitem{Tang:2018rer}
J.~Tang, T.-C. Wang and Y.~Zhang, \emph{{Invisible neutrino decays at the
  MOMENT experiment}},
  \href{https://doi.org/10.1007/JHEP04(2019)004}{\emph{JHEP} {\bfseries 04}
  (2019) 004} [\href{https://arxiv.org/abs/1811.05623}{{\ttfamily
  1811.05623}}].

\bibitem{Abe:2011ks}
{\scshape T2K} collaboration, \emph{{The T2K Experiment}},
  \href{https://doi.org/10.1016/j.nima.2011.06.067}{\emph{Nucl. Instrum. Meth.}
  {\bfseries A659} (2011) 106}
  [\href{https://arxiv.org/abs/arXiv:1106.1238}{{\ttfamily arXiv:1106.1238}}].

\bibitem{Ayres:2007tu}
{\scshape NOvA} collaboration, \emph{{The NOvA Technical Design Report}},
  \href{https://doi.org/10.2172/935497}{\emph{FERMILAB-DESIGN-2007-01} (2007)
  }.

\bibitem{Abe:2015zbg}
{\scshape Hyper-Kamiokande Proto-Collaboration} collaboration, \emph{{Physics
  potential of a long-baseline neutrino oscillation experiment using a J-PARC
  neutrino beam and Hyper-Kamiokande}},
  \href{https://doi.org/10.1093/ptep/ptv061}{\emph{PTEP} {\bfseries 2015}
  (2015) 053C02} [\href{https://arxiv.org/abs/arXiv:1502.05199}{{\ttfamily
  arXiv:1502.05199}}].

\bibitem{Acciarri:2015uup}
{\scshape DUNE} collaboration, \emph{{Long-Baseline Neutrino Facility (LBNF)
  and Deep Underground Neutrino Experiment (DUNE)}},
  \href{https://arxiv.org/abs/arXiv:1512.06148}{{\ttfamily arXiv:1512.06148}}.

\bibitem{Abe:2016ero}
{\scshape Hyper-Kamiokande} collaboration, \emph{{Physics potentials with the
  second Hyper-Kamiokande detector in Korea}},
  \href{https://doi.org/10.1093/ptep/pty044}{\emph{PTEP} {\bfseries 2018}
  (2018) 063C01} [\href{https://arxiv.org/abs/1611.06118}{{\ttfamily
  1611.06118}}].

\bibitem{Asano:2011nj}
K.~Asano and H.~Minakata, \emph{{Large-Theta(13) Perturbation Theory of
  Neutrino Oscillation for Long-Baseline Experiments}},
  \href{https://doi.org/10.1007/JHEP06(2011)022}{\emph{JHEP} {\bfseries 06}
  (2011) 022} [\href{https://arxiv.org/abs/1103.4387}{{\ttfamily 1103.4387}}].

\bibitem{Huber:2004ka}
P.~Huber, M.~Lindner and W.~Winter, \emph{{Simulation of long-baseline neutrino
  oscillation experiments with GLoBES (General Long Baseline Experiment
  Simulator)}}, \href{https://doi.org/10.1016/j.cpc.2005.01.003}{\emph{Comput.
  Phys. Commun.} {\bfseries 167} (2005) 195}
  [\href{https://arxiv.org/abs/hep-ph/0407333}{{\ttfamily hep-ph/0407333}}].

\bibitem{Huber:2007ji}
P.~Huber, J.~Kopp, M.~Lindner, M.~Rolinec and W.~Winter, \emph{{New features in
  the simulation of neutrino oscillation experiments with GLoBES 3.0: General
  Long Baseline Experiment Simulator}},
  \href{https://doi.org/10.1016/j.cpc.2007.05.004}{\emph{Comput. Phys. Commun.}
  {\bfseries 177} (2007) 432}
  [\href{https://arxiv.org/abs/hep-ph/0701187}{{\ttfamily hep-ph/0701187}}].

\bibitem{Agostino:2012fd}
{\scshape MEMPHYS} collaboration, \emph{{Study of the performance of a large
  scale water-Cherenkov detector (MEMPHYS)}},
  \href{https://doi.org/10.1088/1475-7516/2013/01/024}{\emph{JCAP} {\bfseries
  1301} (2013) 024} [\href{https://arxiv.org/abs/1206.6665}{{\ttfamily
  1206.6665}}].

\bibitem{Beacom:2003nk}
J.~F. Beacom and M.~R. Vagins, \emph{{GADZOOKS! Anti-neutrino spectroscopy with
  large water Cherenkov detectors}},
  \href{https://doi.org/10.1103/PhysRevLett.93.171101}{\emph{Phys. Rev. Lett.}
  {\bfseries 93} (2004) 171101}
  [\href{https://arxiv.org/abs/hep-ph/0309300}{{\ttfamily hep-ph/0309300}}].

\bibitem{Campagne:2006yx}
J.-E. Campagne, M.~Maltoni, M.~Mezzetto and T.~Schwetz, \emph{{Physics
  potential of the CERN-MEMPHYS neutrino oscillation project}},
  \href{https://doi.org/10.1088/1126-6708/2007/04/003}{\emph{JHEP} {\bfseries
  04} (2007) 003} [\href{https://arxiv.org/abs/hep-ph/0603172}{{\ttfamily
  hep-ph/0603172}}].

\bibitem{FernandezMartinez:2009hb}
E.~Fernandez-Martinez, \emph{{The gamma = 100 beta-Beam revisited}},
  \href{https://doi.org/10.1016/j.nuclphysb.2010.02.028}{\emph{Nucl. Phys.}
  {\bfseries B833} (2010) 96}
  [\href{https://arxiv.org/abs/0912.3804}{{\ttfamily 0912.3804}}].

\bibitem{Paschos:2001np}
E.~A. Paschos and J.~Y. Yu, \emph{{Neutrino interactions in oscillation
  experiments}}, \href{https://doi.org/10.1103/PhysRevD.65.033002}{\emph{Phys.
  Rev.} {\bfseries D65} (2002) 033002}
  [\href{https://arxiv.org/abs/hep-ph/0107261}{{\ttfamily hep-ph/0107261}}].

\bibitem{Huber:2002mx}
P.~Huber, M.~Lindner and W.~Winter, \emph{Superbeams versus neutrino
  factories}, {\emph{Nucl. Phys.} {\bfseries B645} (2002) 3}
  [\href{https://arxiv.org/abs/hep-ph/0204352}{{\ttfamily hep-ph/0204352}}].

\bibitem{Itow:2001ee}
Y.~Itow et~al., \emph{The jhf-kamioka neutrino project},
  \href{https://arxiv.org/abs/hep-ex/0106019}{{\ttfamily hep-ex/0106019}}.

\bibitem{Ishitsuka:2005qi}
M.~Ishitsuka, T.~Kajita, H.~Minakata and H.~Nunokawa, \emph{Resolving neutrino
  mass hierarchy and cp degeneracy by two identical detectors with different
  baselines}, {\emph{Phys. Rev.} {\bfseries D72} (2005) 033003}
  [\href{https://arxiv.org/abs/hep-ph/0504026}{{\ttfamily hep-ph/0504026}}].

\bibitem{Ambats:2004js}
{\scshape NOvA} collaboration, \emph{Nova proposal to build a 30-kiloton
  off-axis detector to study neutrino oscillations in the fermilab numi
  beamline},  \href{https://arxiv.org/abs/hep-ex/0503053}{{\ttfamily
  hep-ex/0503053}}.

\bibitem{Yang_2004}
{\scshape NOvA} collaboration, \emph{Study of physics sensitivity of $\nu_\mu$
  disappearance in a totally active version of nova detector},
  \href{https://arxiv.org/abs/Off-Axis-Note-SIM-30}{{\ttfamily
  Off-Axis-Note-SIM-30}}.

\bibitem{GLoBES}
``{GLoBES (2018)}.'' \url{https://www.mpi-hd.mpg.de/personalhomes/globes/}.

\bibitem{Alion:2016uaj}
{\scshape DUNE} collaboration, \emph{{Experiment Simulation Configurations Used
  in DUNE CDR}},  \href{https://arxiv.org/abs/arXiv:1606.09550}{{\ttfamily
  arXiv:1606.09550}}.

\bibitem{Hyper-Kamiokande:2016dsw}
{\scshape Hyper-Kamiokande} collaboration, \emph{{Hyper-Kamiokande Design
  Report}}, .

\bibitem{Acciarri:2016crz}
{\scshape DUNE} collaboration, \emph{{Long-Baseline Neutrino Facility (LBNF)
  and Deep Underground Neutrino Experiment (DUNE)}},
  \href{https://arxiv.org/abs/1601.05471}{{\ttfamily 1601.05471}}.

\bibitem{Strait:2016mof}
{\scshape DUNE} collaboration, \emph{{Long-Baseline Neutrino Facility (LBNF)
  and Deep Underground Neutrino Experiment (DUNE)}},
  \href{https://arxiv.org/abs/1601.05823}{{\ttfamily 1601.05823}}.

\bibitem{Acciarri:2016ooe}
{\scshape DUNE} collaboration, \emph{{Long-Baseline Neutrino Facility (LBNF)
  and Deep Underground Neutrino Experiment (DUNE)}},
  \href{https://arxiv.org/abs/1601.02984}{{\ttfamily 1601.02984}}.

\end{thebibliography}\endgroup

\end{document}